\documentstyle[12pt,epsfig]{article}
 1
 1
 1

\begin{document}
\bibliographystyle{unsrt}

\def\lta{\;\raisebox{-.5ex}{\rlap{$\sim$}} \raisebox{.5ex}{$<$}\;}
\def\gta{\;\raisebox{-.5ex}{\rlap{$\sim$}} \raisebox{.5ex}{$>$}\;}
\def\grle{\;\raisebox{-.5ex}{\rlap{$<$}}    \raisebox{.5ex}{$>$}\;}
\def\legr{\;\raisebox{-.5ex}{\rlap{$>$}}    \raisebox{.5ex}{$<$}\;}

\newcommand{\beq}{\begin{equation}} 
\newcommand{\eeq}{\end{equation}}
\newcommand{\bea}{\begin{eqnarray}} 
\newcommand{\eea}{\end{eqnarray}}
\newcommand{\nl}{\nonumber \\} 
\newcommand{\np}{Nucl.\,Phys.\,}
\newcommand{\pl}{Phys.\,Lett.\,}
\newcommand{\pr}{Phys.\,Rev.\,}
\newcommand{\prl}{Phys.\,Rev.\,Lett.\,}
\newcommand{\prep}{Phys.\,Rep.\,}
\newcommand{\zp}{Z.\,Phys.\,}
\newcommand{\sovjnp}{{\em Sov.\ J.\ Nucl.\ Phys.\ }}
\newcommand{\nuclinst}{{\em Nucl.\ Instrum.\ Meth.\ }}
\newcommand{\annp}{{\em Ann.\ Phys.\ }}
\newcommand{\intjmp}{{\em Int.\ J.\ of Mod.\  Phys.\ }}
\newcommand{\ra}{\rightarrow}
\newcommand{\permille}{$^0 \!\!\!\: / \! _{00}$}
\newcommand{\dd}{{\rm d}}
\newcommand{\oal}{{\cal O}(\alpha)}%
\newcommand{\su}{$ SU(2) \times U(1)\,$}
\newcommand{\xdim}{$dim=6$~} 
\newcommand{\eps}{\epsilon}
\newcommand{\mw}{M_{W}}
\newcommand{\mww}{M_{W}^{2}}
\newcommand{\mbb}{m_{b \bar b}}
\newcommand{\mcc}{m_{c \bar c}}
\newcommand{\mbc}{m_{b\bar b(c \bar c)}}
\newcommand{\mh}{m_{H}}
\newcommand{\mhh}{m_{H}^2}
\newcommand{\mz}{M_{Z}}
\newcommand{\mzz}{M_{Z}^{2}}
\newcommand{\lra}{\leftrightarrow}
\newcommand{\tr}{{\rm Tr}}
\newcommand{\ie}{{\em i.e.}}
\newcommand{\cm}{{{\cal M}}}
\newcommand{\cl}{{{\cal L}}}
\def\Ww{{\mbox{\boldmath $W$}}}  
\def\B{{\mbox{\boldmath $B$}}}         
\def\nn{\noindent}
\newcommand{\sinsq}{\sin^2\theta}
\newcommand{\cossq}{\cos^2\theta}
\newcommand{\epem}{$e^{+} e^{-}\;$}
\newcommand{\epemt}{e^{+} e^{-}\;}
\newcommand{\eeah}{$e^{+} e^{-} \ra H \gamma \;$}
\newcommand{\eahnw}{$e\gamma \ra H \nu_e W$}
\newcommand{\thebb}{\theta_{b-beam}}
\newcommand{\thebc}{\theta_{b(c)-beam}}
\newcommand{\pte}{p^e_T}
\newcommand{\ptH}{p^H_T}
\newcommand{\gag}{$\gamma \gamma$ }
\newcommand{\gam}{\gamma \gamma }
\newcommand{\aatoh}{$\gamma \gamma \ra H \;$}
\newcommand{\egam}{$e \gamma \;$}
\newcommand{\eat}{e \gamma \;}
\newcommand{\eaeh}{$e \gamma \ra e H\;$}
\newcommand{\eaehb}{$e \gamma \ra e H \ra e (b \bar b)\;$}
\newcommand{\egebb}{$e \gamma (g) \ra e b \bar b\;$}
\newcommand{\egecc}{$e \gamma (g) \ra e c \bar c\;$}
\newcommand{\egebc}{$e \gamma (g) \ra e b \bar b(e c \bar c)\;$}
\newcommand{\eaebb}{$e \gamma \ra e b \bar b\;$}
\newcommand{\eaecc}{$e \gamma \ra e c \bar c\;$}
\newcommand{\aah}{$\gamma \gamma H\;$}
\newcommand{\zah}{$Z \gamma H\;$}
\newcommand{\zzh}{$Z Z H\;$}
\newcommand{\pe}{P_e}
\newcommand{\pg}{P_{\gamma}}
\newcommand{\delbb}{\Delta m_{b \bar b}}
\newcommand{\delbc}{\Delta m_{b \bar b(c\bar c)}}
\renewcommand{\d}{{\rm d}}
\newcommand{\db}{{\rm d}_{\scriptscriptstyle{\rm B}}}
\newcommand{\bard}{\overline{{\rm d}}}
\newcommand{\bardb}{\overline{{\rm d}}_{\scriptscriptstyle{\rm B}}}
\renewcommand{\aa}{{\rm d}_{\scriptscriptstyle \gamma\gamma}}
\newcommand{\GeV}{\rm GeV}
\newcommand{\az}{{\rm d}_{\scriptscriptstyle \gamma Z}}
\newcommand{\zz}{{\rm d}_{\scriptscriptstyle ZZ}}
\newcommand{\aacp}{\overline{{\rm d}}_{\scriptscriptstyle \gamma\gamma}}
\newcommand{\azcp}{\overline{{\rm d}}_{\scriptscriptstyle \gamma Z}}
\newcommand{\zzcp}{\overline{{\rm d}}_{\scriptscriptstyle ZZ}}
\newcommand{\tgw}{\tan{\theta_W}}
\newcommand{\cotgw}{(\tan{\theta_W})^{-1}}

\begin{titlepage}
\vskip 22pt 

\noindent
\begin{center}
{\Large \bf Looking for  anomalous 
{\boldmath \aah} and {\boldmath \zah} couplings \\ 
at future linear colliders }
\end{center}
\bigskip

\begin{center}
{\large 
E.~Gabrielli~$^{a},$
$\;\,$V.A.~Ilyin~$^b \;\,$
and $\;\,$B.~Mele~$^c$. 
} \\
\end{center}

\medskip\noindent
$^a$ Institute of Theoretical Physics, C-XVI,
University Aut\'onoma of Madrid, Spain \\
\noindent
$^b$ Institute of Nuclear Physics, Moscow State University, Russia \\
\noindent
$^c$ INFN, Sezione di Roma 1 and Rome University ``La Sapienza", Italy
\bigskip
\begin{center}
{\bf Abstract} \\
\end{center}
{\small 
We consider the possibility of studying anomalous contributions to the
\aah and \zah vertices through the process \eaeh
at future $e \gamma$ linear colliders, with $\sqrt{S}=(500-1500)$~GeV. 
We make a model independent analysis based on \su invariant
effective operators of \xdim added to the standard model lagrangian.
We consider a  light Higgs boson (mostly decaying in $b\bar b$ pairs),
and include all the relevant backgrounds. Initial $e$-beam
polarization effects are also 
analyzed.
We find that the process \eaeh provides an excellent opportunity
to  strongly constraint
both the CP-even and the CP-odd anomalous  
contributions to the \aah and \zah vertices.
}
\vskip 11pt 
\vfil

\vskip 22pt 
\vfil
\noindent
e-mail: \\
emidio.gabrielli@cern.ch, $\;$
ilyin@theory.npi.msu.su, $\;$
mele@roma1.infn.it
\end{titlepage}

\section{Introduction}

Higgs boson physics is a major aim in the program of present and future
colliders.
Presently, direct experimental mass limits on $\mh$ come from  
searches at LEP2, giving $\mh \gta \mz$. Also, LEP1 precision
measurements consistency  sets an upper limit
$\mh \lta 300$~GeV in the SM \cite{dago}.

Once the Higgs boson will be discovered, establishing  its properties and 
interactions will be crucial either to consolidate the standard model (SM)
or to find possible anomalies.
Indeed, the symmetry breaking sector of the theory is expected to be 
particularly sensitive to the presence of possible new physics (NP) effects 
that could eventually even explain its origin.

The interactions of the Higgs boson  with
the neutral electroweak gauge bosons,  $\gamma$ and $Z$, are particularly
interesting. 
By measuring the three vertices $ZZH$, $\gamma\gamma H$ and $Z\gamma H$, 
one can test a delicate feature of the Standard Model, that is the 
relation between the spontaneous symmetry-breaking
mechanism and the electroweak mixing of the two gauge groups $SU(2)$ and
$U(1)$.   
While  the $ZZH$ vertex occurs at the
tree level, the other two contribute only at one loop
in the SM. As a consequence the \aah and \zah couplings are particularly 
sensitive to possible contributions from new physics, such as the existence 
of new heavy particles circulating in the loop.

A measurement of the \aah and \zah couplings could be performed
at LHC, by determining the branching ratio ($B$) of the corresponding 
Higgs boson decays $H\ra \gamma \gamma$ and $H\ra Z \gamma$. This holds 
for a rather light Higgs, i.e. $m_H\lta 140$~GeV [($m_H\gta 115$~GeV 
is also required in the latter channel, which guarantees a 
$B(H\ra Z \gamma)$ as large as ${\cal O}(10^{-3})$]. 
Measuring the \zah vertex is anyhow more complicated.  
The $H\ra Z \gamma$ final states include the $Z$ decay products, 
either jets or lepton pairs, for which much heavier backgrounds are expected. 

A more accurate determination of both the \aah and \zah vertices
is expected at future linear colliders with c.m. energy 
$\sqrt{s}\simeq (300\div 2000)$~GeV
and integrated luminosity ${\cal O}(100 \div 1000)$ fb$^{-1}$
\cite{saar,zerw}.
Besides \epem collisions, linear colliders give the possibility,
through laser backscattering \cite{spec,mono},
to study $\gamma\gamma$ and $\gamma e$ interactions at energies and
luminosities comparable to those in \epem collisions.
While the process $\gamma\gamma \to H$ on the Higgs boson resonance
will be the ideal place where to measure the \aah coupling,
the reaction \eaeh will give a unique possibility to study both
the \aah (without requiring a fine tuning of the c.m. energy,
as for the process $\gamma\gamma \to H$) and \zah vertices with good 
statistics. 
In fact, the crossed channel
$ e^+e^- \to \gamma H$ suffers from very low cross sections 
and overwhelming backgrounds \cite{barr,abba,djou}.
On the other hand, in the channel $e^+e^- \to Z H$
also involving the \zah coupling, the latter contributes to the
corresponding  one-loop corrections in the SM, 
thus implying a large tree-level background over which to study
possible small deviations.

In the SM, the one-loop process   $e\gamma \to eH$
was analysed in details  \cite{noii}, stressing the unique possibility
offered by this channel to study the \zah coupling in a clear 
experimental environment and with good statistics.
In particular in \cite{noii}, 
we remarked how requiring large transverse momentum
events automatically selects more efficiently the \zah 
contributions with respect to the \aah interaction that is dominant in the
total cross section.

In this paper we analyze the prospects of the \eaeh reaction in setting
experimental bounds on the value of the anomalous \aah and \zah couplings. 
Some preliminary results have been presented in 
\cite{prel}\footnote{Note that in this talk the factor $(-g^\sigma_e)$
(the $Z$ charge of electron) is missing in the formula for the anomalous
contribution to the amplitude.}.
We adopt a model independent approach, 
where an effective lagrangian is obtained by adding
\xdim \su invariant operators  to the SM Lagrangian. 
In realistic  models extending the SM, these operators contribute in some 
definite combinations. 
However, this approach can give some general  insight into the
problem when discussing  possible deviations from  the standard-model
one-loop Higgs vertices.  
The anomalous operators we will consider contribute to 
the three vertices  \aah, \zah and $ZZH$,  only the first two 
being involved in the \eaeh reaction.  
Although the anomalous contributions to the \aah vertex 
can be more efficiently bounded through the resonant $\gamma\gamma\to H$ 
reaction, independent interesting bounds can be obtained by measuring 
the total rate of the \eaeh process. On the other hand, the latter channel
offers a unique possibility to bound anomalous \zah couplings too,
the latter being enhanced in large $p_T$ events.
Some general analysis  of the potential of the two processes above in 
bounding the \aah and \zah vertices has been given in \cite{ginz}.
Here we provide a more detailed study of the process
\eaeh at c.m. energies of 500 and 1500~GeV,
when the Higgs boson decays predominantly in $b$-quark pairs,
(i.e. for $\mh \lta 140$~GeV) that includes all
the relevant backgrounds and initial beam polarization effects.

We stress that in our analysis we assume a monochromatic photon beam. 
Presently\footnote{We acknowledge thorough and enlightening 
discussions with V.~Telnov on this point.}, experts on
photon-beams construction in $e^+e^-$ colliders
claim that assuming a particular form for the photon beam energy spectrum
(as is often done in the literature) is somewhat  premature.
According to what it is actually known, it seems {\it cleaner}, 
as far as predictions are concerned, to simply assume
a monochromatic photon beam with an energy equal to the expected peak
energy of the laser backscattered photons, which is about 0.8 times  
the energy of the basic electron beam.
This, for instance, gives \egam collisions with $\sqrt{s}= 0.9$ TeV,
when starting from a $e^+e^-$ colliders at $\sqrt{s}= 1$ TeV.
Regarding the luminosity of the \egam machine, 
this is presently expected to be comparable with the luminosity of the
initial $e^+e^-$ machine (the most pessimistic assumptions
gives a reduction factor of about 1/3 with respect to the  
corresponding $e^+e^-$ machine).

The paper is organized as follows.
In  section 2, we summarize the \eaeh process main features
in the SM. We then review the channels that give important
backgrounds for the 
$e b\bar b$  final state.
In section 3, we introduce the effective lagrangian 
describing the relevant anomalous interactions and give
the analytical expressions for the corresponding helicity amplitudes 
for the process \eaeh. 
In section 4 we discuss our strategy for getting bounds on the anomalous 
\aah and \zah couplings from \eaeh.
We describe the  kinematical cuts that optimize the signal to background
ratio ($S/B$), including the SM signal among the ``backgrounds". 
Sections 5 contains our numerical results.
In section 6, we discuss the resulting limits that can be put on the 
New Physics (NP) scale $\Lambda$ and give our conclusions
\footnote{Most of the results presented in this work were obtained 
with the help 
of the CompHEP package \cite{comp}.}. 

\section{The reaction \eaeh in the SM and other background channels}

In this section we describe the main channels that contribute as  backgrounds
in the search for anomalous couplings through the process \eaeh
followed by the Higgs boson decay $H\to b \bar b$.
We first summarize the main feature of the \eaeh process
in the SM, that, in a sense, is the first ``irreducible"
background to be taken into account when searching for an anomalous signal.

In \cite{noii}, we presented the complete analytical results for the helicity 
amplitudes of the \eaeh process (see also \cite{cott}). 
This amplitude is given in terms of the contributions denoted as 
`\aah' and `\zah', which are related to the \aah and \zah vertices 
respectively, and a 'BOX' contribution. 
The separation of the rate into these three parts corresponds 
to the case  where the Slavnov-Taylor identities for the `\aah' 
and `\zah' Green  functions just imply the transversality with respect to 
the incoming photon momentum (see also \cite{ginz} for a detailed discussion
of this formally delicate separation). 

The total rate of this reaction is rather large. In particular, if
$\sqrt{s}\gta 500$~GeV, one finds $\sigma>10$ fb
for $\mh$ up to about 250~GeV, 
and  $\sigma>1$ fb for $\mh$ up to about 400~GeV. 
In the total cross section,
the main contribution is given by the \aah vertex because of
the massless photon propagator in the $t$-channel. 
On the other hand, the \zah vertex contribution 
is depleted by the $Z$ propagator in the $t$-channel. 
As discussed in \cite{noii} and \cite{ginz}, 
the \zah vertex effects can then be 
enhanced by requiring a final electron tagged at large angle.
Indeed, the latter kinematical configuration 
strongly depletes the photon propagator and, hence, 
the \aah vertex contribution, while keeping most of the \zah contribution.
For example, for $\pte>100$~GeV and $\sqrt{s}\simeq 500$~GeV, 
we found that the \zah contribution to the cross section is about 60\% 
of the \aah one, hence giving a considerable fraction 
of the  production rate, that, at the same time, is still sufficient 
to guarantee investigation (about 0.7 fb for $\mh\sim 120$~GeV).

In what follow, we will consider also the possibility of having 
longitudinally polarized electron beams. In \cite{noii}, we studied, 
 for $m_H=120$~GeV, the $e$-beam polarization dependence of  the SM total
cross section and of its \aah, \zah and BOX components, 
and  their  interference 
pattern.  The latter turns out to be particularly
sensitive to the electron polarization. For instance, assuming 
left-handed electrons with $\pe=-1$
(right-handed electrons with $\pe=+1$)  the total cross section 
increases (decreases) by about 94\% at $\sqrt{s}=500$~GeV. 
Indeed, for $\pe=-1$ ($\pe=+1$), there is a strong 
constructive (destructive) interference
between the \aah and \zah contributions to the amplitude.

Apart from the SM signal in \eaeh, 
the relevant backgrounds in our problem are all the background channels
analyzed in \cite{noii} for the SM  $e\gamma\to eH\to eb\bar b$ process.
The main irreducible background 
comes from the direct production of $b$-quark pairs in \eaebb.  
An angular cut on the $b$'s with respect to both the
beams can reduce the rate for 
this potentially dangerous background to a comparable level
with the SM signal.
Also, whenever  the $c$ quarks are misidentified into $b$'s,
a further source of background is given by the charm-pair
production through \eaecc.
Assuming about 10\% of probability  of misidentifying a  $c$ quark 
into a $b$, the contribution of  the \eaecc ``effective rate" 
turns out to be of the same order as the \eaebb rate. 

A further background, analyzed in \cite{noii}, is the resolved \egebc
production, where the photon interacts via its gluonic content. 
This was found to be negligible  with respect to the previous channels.

At moderate values of $\sqrt{s}$ ($\sqrt{s}\sim 500$~GeV), 
further improvements in the $S/B$ ratio can be obtained by exploiting the
final-electron angular asymmetry in the SM signal \cite{noii}. Indeed,  
the final electron in \eaeh moves mostly in the forward direction.   
On the other hand, for the \eaebb background 
the final electron angular distribution, although not
completely symmetric, is almost equally shared in the forward and backward
direction with respect to the beam. 
On the other hand, for $\sqrt{s}\gta 1$~TeV, a clear asymmetry develops in the
background channel too. In fact,
the contribution of the diagrams with an
electron exchanged in the u-channel (responsible for the backscattering of
final electrons) decreases as $1/s$. Hence, at large $\sqrt{s}$, 
one does not get improvements in the S/B
ratio by exploiting this forward-backward asymmetry. 
Anyhow, even at lower values of $s$,
although the $S/B$ ratio can be  improved 
up to $\sim 1$, the bounds discussed below  depend only marginally 
on the electron angular cut.

After a detailed study of the SM signal versus all the relevant backgrounds,
in \cite{noii} we concluded that with a
luminosity of 100 fb$^{-1}$, at $\sqrt{s}=500$~GeV, one expects an accuracy as
good as about 10\% on the measurement of the \zah effects, assuming the 
validity of the SM. Here, we will use a similar strategy 
to study the sensitivity of the \eaeh process to 
possible anomalous coupling contributions in both the \aah and \zah vertices.

\section{The effective Lagrangian}

In our study, we want to set the sensitivity of the \eaeh process
to possible anomalies in the \aah and \zah interactions
in a most general framework.
To this end, we adopt the effective Lagrangian formalism \cite{anomOP}
that is able to describe the low-energy residual effects of some complete 
theory correctly describing NP at higher energy scales. 
This introduces 
\xdim $SU(2)\times U(1)$  invariant operators.
Among the different operators involving the Higgs boson,
the ones contributing to the \eaeh amplitude via anomalous couplings 
in the \aah and \zah vertices are particularly interesting here.
We then will concentrate on the so-called ``superblind" operators,
that is the ones 
not leading at the same time to anomalous gauge-boson self couplings
that can be more comfortably constrained in processes like 
$e^+e^- \to WW$.
In particular, assuming the so-called custodial symmetry, two pairs
(the first CP-even and the second CP-odd) of effective operators 
are relevant for our problem

\beq
{\cal L}^{eff} = \d\cdot {\cal O}_{UW} + \db\cdot {\cal O}_{UB} +
     \bard\cdot \bar {\cal O}_{UW} 
                        + \bardb\cdot \bar {\cal O}_{UB},
\eeq

\beq
{\cal O}_{UW} \,=\, \frac{1}{v^2} \left( |\Phi|^2-\frac{v^2}{2}\right)
    \cdot W^{i\mu\nu}  W^i_{\mu\nu}, \qquad
   {\cal O}_{UB} \,=\, \frac{1}{v^2} \left( |\Phi|^2-\frac{v^2}{2}\right)
    \cdot B^{\mu\nu}  B_{\mu\nu},
\eeq

\beq
\bar {\cal O}_{UW} \,=\, \frac{1}{v^2}  |\Phi|^2
    \cdot W^{i\mu\nu}  \tilde W^i_{\mu\nu}, \qquad
   \bar {\cal O}_{UB} \,=\, \frac{1}{v^2} |\Phi|^2
    \cdot B^{\mu\nu}  \tilde B_{\mu\nu}, 
\eeq
where
$ \tilde W^i_{\mu\nu} = \epsilon_{\mu\nu\mu'\nu'}\cdot W^{i\mu'\nu'}$ and
$ \tilde B_{\mu\nu} = \epsilon_{\mu\nu\mu'\nu'}\cdot B^{\mu'\nu'}$.
In these formulas $\Phi$ is the Higgs doublet and 
$v$ is the electroweak vacuum expectation value.

Accordingly, the \aah and \zah anomalous terms contributions to 
the helicity amplitudes of \eaeh are given by:
\beq
M_{anom}(\sigma,\lambda) = 
        M^{\gamma\gamma}(\sigma,\lambda) +
        M^{\gamma Z}(\sigma,\lambda),
\eeq
where
\bea
M^{\gamma\gamma}(\sigma,\lambda) &=&
     \frac{4\pi\alpha}{M_Z (-t)} \sqrt{-\frac{t}{2}}
    \{ d_{\gamma\gamma} [(u-s)-\sigma\lambda(u+s) 
      -i \bar d_{\gamma\gamma} [\lambda (u-s)+\sigma (u+s)]\}, \nl
M^{\gamma Z}(\sigma,\lambda) &=&
     \frac{4\pi\alpha (-g_e^\sigma)}{M_Z (M_Z^2-t)} \sqrt{-\frac{t}{2}}
    \{ d_{\gamma Z} [(u-s)-\sigma\lambda(u+s) 
      -i \bar d_{\gamma Z} [\lambda (u-s)+\sigma (u+s)]\}.
\eea  
Here, $s$, $t$ and $u$ are the Mandelstam kinematical variables
(defined as in \cite{noii}), $\sigma/2=\pm 1/2$
and $\lambda=\pm 1$ are the electron and photon  helicities, respectively. 
The $Z$ charge of electron is denoted as $g_e^\sigma$.
The anomalous couplings $\d, \db, \bard, \bardb$ contribute 
to the \aah, \zzh, and \zah interactions in the combinations 
\bea
\aa~&=&~ \tgw~\d~+~\cotgw~\db,~~~\az~=~\d~-~\db\nl
\zz~&=&~ \cotgw~\d~+~\tgw~\db,\nl
\aacp~&=&~ \tgw~\bard~+~\cotgw~\bardb,~~~\azcp~=~\bard~-~\bardb\nl
\zzcp~&=&~ \cotgw~\bard~+~\tgw~\bardb,
\label{eq:anomC}
\eea
where $\theta_W$ is the Weinberg angle.
We assume 
$\sin^2\theta_W=0.2247$
\footnote{All the physical parameters of the SM used here
are the same as in our paper\cite{noii}.}.
Note that there is no interference between the CP-odd terms 
(with couplings $\bar d$)
and any CP-even triangle diagram  inducing the \aah and \zah couplings 
in the SM amplitude, although the interference  with 
the SM box amplitude is nonvanishing.  
This is due to the real value of the SM amplitude induced by the
triangle diagrams  for $M_H<2M_W$ and $M_H<2m_{top}$ (one can neglect
contributions of light quark loops), while the box diagrams contribute with
complex numbers. Since  the box contribution to 
the SM amplitude is rather small,
the complete  cross section 
(i.e., the one deriving from the SM Lagrangian and the  effective one)
depends  only marginally on the sign of the CP-odd
couplings $\aacp$ and $\azcp$. On the contrary, for the CP-even terms the
interference is relevant,  and the dependence on the $\aa$ and $\az$ coupling 
is not symmetrical with  respect to the corresponding SM point 
($\aa=0$, $\az=0$). This effect,
generally speaking, decreases the sensitivity to the CP violating anomalous
couplings.

\section{Bounds on anomalous \aah and \zah couplings}

In this section we develop a strategy to work out the sensitivity
of the process \eaeh to the anomalous couplings
$\aa,~\az$ and $\aacp,~\azcp$ introduced by the effective lagrangian
defined in the previous section.
The ability to distinguish an anomalous signal will depend on
the relative importance of the deviation in the 
\eaeh cross section $\sigma_S(\kappa)$, where 
\beq
   \kappa = \aa , \az, \aacp, \azcp,           
\eeq
over the SM background. The latter will be made up of both the
SM \eaeh cross section $\sigma_S(0)$ and all the relevant
backgrounds $\sigma_B$ to the $e\gamma\to eH\to eb\bar b$ signal.
As discussed in Section 2, $\sigma_B$ is given by the 
processes \eaebb and \eaecc (reducing  the \eaecc cross section  
by a factor 1/10 in order to take into account a 10\%
probability of misidentifying a $c$ quark into a $b$ quark).
In this comparison, the available total integrated luminosity
${\cal L}_{int}$ is of course a crucial parameter.
In particular, we compare the excess in the number of observed events
due to anomalous couplings
\beq
   N^{\rm anom}(\kappa) = {\cal L}_{int} \cdot 
     [\sigma_S(\kappa)-\sigma_S(0)]\;
\eeq
with the total number of observed events
\beq
N^{\rm tot}(\kappa) = {\cal L}_{int} \cdot [\sigma_S(\kappa)+\sigma_B]\;.
\eeq
Then, the requirement that no deviation from the SM cross section is 
observed at the 95\% CL reads:
\beq
N^{\rm anom}(\kappa) < 1.96 \cdot \sqrt{N^{\rm tot}(\kappa)}.
\label{duesig}
\eeq
In the following, we assume $\mh=120$~GeV and make the analysis at two
different c.m. collision energies, i.e. 
$\sqrt{s}=500$~GeV and 1.5~TeV, with integrated luminosity
${\cal L}_{int}=10^2$ and $10^3$ fb$^{-1}$, respectively.
Correspondingly, the cross sections $\sigma_S(\kappa)$ 
and $\sigma_B$ will be subjected to two different sets of kinematical cuts,
optimizing the $S/B$ ratio according to the c.m. collision energy :

a) at $\sqrt{s}=500$ GeV, we require $\thebc>18^o$

b) at $\sqrt{s}=1500$ GeV, we require $\thebc>8^o$

\noindent
where $\thebc$ stands for the angle that each final $b(c)$ quark forms with
either of the initial beams. These cuts strongly deplete the
\eaebb and \eaecc backgrounds. Also, we  integrate
the invariant mass distribution of the $b(c)$ pairs only on the interval
$\pm \Delta \mbc =\pm 3$~GeV around $\mh$, in order to keep only
that fraction of the background that can fake the Higgs boson decay
within the experimental resolution.

Furthermore, in order to enhance either the $\aa$ or the $\az$ vertex 
sensitivity,  we either integrate over the 
complete final electron transverse 
momentum $\pte$ or put a minimum cut $\pte>100$~GeV.

At $\sqrt{s}=500$~GeV, in the $\pte>100$~GeV case we apply a further cut
$\theta_e <90^o$, that is we select only the events with a forward final 
electrons. In this way, we exploit the stronger electron asymmetry in the
\eaeh process (both in its SM and its anomalous components)
with respect to the \eaebb and \eaecc backgrounds.
At larger $\sqrt{s}$ this difference tends to fade away. 
\begin{table}
\begin{center}
\begin{tabular}{|r||c|c|c|c|c|c|c|}
\hline 
$(\pe,~\pte )$
& (0,0)  &  (0,100) & (1,0)  & (1,100) & (-1,0) & (-1,100)
\\ \hline 
   \hline $\sigma_B$ & $ 7.69 $ & $0.432 $ & 
                  $ 6.39 $ & $ 0.230 $ & 
                  $ 9.02 $ & $ 0.632 $ 
\\  \hline $\sigma_S(0)$ & $ 3.26$ & $ 0.370$ & 
                  $ 2.66$ & $ 0.0264$ & 
                  $ 3.86$ & $ 0.718$ 
\\   \hline $C_1\times 10^{-2}$ & $-14.7$ & $-1.55$ & 
                  $-12.8$ & $-0.462$ & 
                  $-16.4$ & $-2.66$ 
\\ \hline $C_2\times 10^{-1}$  & $-8.47$ & $-5.94$ & 
                  $ 4.17$ & $ 1.79$ & 
                  $-21.1$ & $-13.6$ 
\\ \hline $C_3$  & $-4.50$ & $-2.30$ & 
                  $ 0.50$ & $-0.30$ & 
                  $-1.0$ & $-5.50$ 
\\ \hline $C_4$  & $-2.25$ & $-1.32$ & 
                  $-0.450$ & $-0.175$ & 
                  $-3.95$ & $-2.85$ 
\\ \hline $C_5\times 10^{-4}$  & $ 17.8$ & $ 2.57$ & 
                  $ 17.7$ & $ 2.58$ & 
                  $ 17.7$ & $ 2.57$ 
\\ \hline $C_6\times 10^{-3}$  & $ 8.49$ & $ 5.96$ & 
                  $ 6.84$ & $ 4.74$ & 
                  $ 10.2$ & $ 7.03 $ 
\\ \hline $C_7\times 10^{-4}$  & $ 17.7$ & $ 2.58$ & 
                  $ 17.8$ & $ 2.58$ & 
                  $ 17.7$ & $ 2.57$ 
\\ \hline $C_8\times 10^{-3}$  & $ 8.62$ & $ 5.95$ & 
                  $ 6.93$ & $ 4.76$ & 
                  $ 10.1$ & $ 7.06$ 
\\ \hline $C_9\times 10^{-3}$  & $ 3.73$ & $ 2.48$ & 
                  $-34.5$ & $-21.9$ & 
                  $ 42.3$ & $ 27.0$ 
\\ \hline $C_{10}\times 10^{-3}$ & $ 3.81$ & $ 2.52$ & 
                  $-35.9$ & $-21.8$ & 
                  $ 41.9$ & $ 26.9$ 
\\ \hline
\end{tabular}
\caption[]{$\sigma_B$, $\sigma_S(0)$ and $C_i$ coefficients 
(in $\mbox{fb}$) for
$\sqrt{S}=500$~GeV, e-beam polarizations
$\pe=0,1,-1$ and final electron transverse momentum cuts 
$\pte > 0$ and  $\pte > 100$~GeV (a cut on
the final electron angle $\theta_e < 90^o$ is applied 
for $\pte > 100$~GeV).}
\label{tab500_ctp}
\end{center}
\end{table}
\begin{table}
\begin{center}
\begin{tabular}{|r||c|c|c|c|c|c|c|}
\hline 
$(\pe,~\pte )$
& (0,0)  &  (0,100) & (1,0)  & (1,100) & (-1,0) & (-1,100)
\\ \hline 
   \hline $\sigma_B$ & $ 2.48 $ & $ 0.398 $ & 
                  $ 2.18 $ & $ 0.210 $ & 
                  $ 2.78 $ & $ 0.586 $ 
\\   \hline $\sigma_S(0)$  & $ 1.57$ & $ 0.437$ & 
                   $ 1.05$ & $ 0.0262$ & 
                   $ 2.11$ & $ 0.838$ 
\\   \hline $C_1\times 10^{-2}$  & $-7.44$ & $-2.25$ & 
                   $-5.29$ & $-0.454$ & 
                   $-9.57$ & $-4.04$ 
\\ \hline $C_2\times 10^{-2}$  & $-1.12$ & $-1.02$ & 
                   $ 0.280$ & $ 0.175$ & 
                   $-2.55$ & $-2.23$ 
\\ \hline $C_3$  & $ 0$ & $ 0.20$ & 
                   $-4.50$ & $ 0.150$ & 
                   $-3.50$ & $-0.650$ 
\\ \hline $C_4$ &  $-0.60$  & $-0.70$ & 
                   $-0.150$ & $-0.150$ & 
                   $-0.650$ & $-0.60$ 
\\ \hline $C_5\times 10^{-4}$  & $ 12.2$ & $  5.79$ & 
                   $ 12.2$  & $ 5.79$ & 
                   $ 12.2$  & $ 5.83$ 
\\ \hline $C_6\times 10^{-4}$   & $ 1.92$ & $ 1.68$ & 
                   $ 1.53$ & $ 1.34$ & 
                   $ 2.31$ & $ 2.03$ 
\\ \hline $C_7\times 10^{-4}$  & $ 12.2$ & $ 5.77$ & 
                   $  12.2$ & $ 5.80$ & 
                   $  12.2$ & $ 5.80$ 
\\ \hline $C_8\times 10^{-4}$  & $ 1.94$ & $ 1.68$ & 
                   $ 1.54$ & $ 1.34$ & 
                   $ 2.28$ & $ 2.02$ 
\\ \hline $C_9\times 10^{-4}$  & $ 0.739$ & $ 0.663$ & 
                   $-6.51$ & $-5.53$ & 
                   $ 8.05$ & $ 6.79$ 
\\ \hline $C_{10}\times 10^{-4}$  & $ 0.741$ & $ 0.639$ & 
                   $-6.43$ & $-5.48$ & 
                   $ 7.90$ & $ 6.84$ 
\\ \hline
\end{tabular}
\caption[]{ The same results as in table \ref{tab500_ctp}, 
but for $\sqrt{S}=1500$~GeV (with no restriction on the final
electron angle for  $\pte > 100$~GeV).} 
\label{tab1500_ctp}
\end{center}
\end{table}

After applying all the cuts described above, the cross section
$\sigma_S(\kappa)$ can be parameterized as a quadratic form in the couplings
$\kappa$
\bea
\sigma_S(\kappa)&=&\sigma_S(0)+\aa C_1+\az C_2+\aacp C_3+
\azcp C_4+\aa^2 C_5+\az^2 C_6+\aacp^2 C_7+\azcp^2 C_8\nonumber\\
&+&\aa\az C_9+\aacp\azcp C_{10}
\label{param}
\eea
where the numerical coefficients for $\sigma_S(0)$ and
$C_i$   for $\mh=120$~GeV  are given in tables 1 and 2
for $\sqrt{S}=500$~GeV and $\sqrt{S}=1500$~GeV, respectively.
The corresponding values of $\sigma_B$ are also shown.
With the help of the above parameterization and according to the
definition of $N^{\rm anom}(\kappa)$
and    $N^{\rm tot}(\kappa)$ given above, one can than develop all 
 the numerical analysis that we are going to present 
in the next section.

\section{Numerical results}
We now present the results of our analysis. In figs. 1-16, 
for two values of $\sqrt{s}$ (500 and  1500~GeV), 
and of the minimum cut on $\pte$  (0 and 100~GeV),
we show the contour plots that
select the area that satisfies the condition in eq.(\ref{duesig}), 
in the planes of some interesting combination of pairs of couplings
[i.e., $(\aa,~\az)$, $(\aa,~\aacp)$, $(\az,~\azcp)$, $(\aacp,~\azcp)$].
All the couplings different from the ones on the
$(x,y)$ axis are assumed to be zero. The solid, dot-dashed, dashed 
(generally closed) lines refer to the values $\pe=0,-1,+1$
for the polarization of the $e$ beam, respectively.
The darker curve among the three (or two) curves for each case
corresponds to the set of points where the cross section gets its SM 
value, that is the set of solutions $\kappa_{SM}$ of the equation
\beq
 \sigma_S(\kappa_{SM})=\sigma_S(0) \;\; \to \;\; 
N^{\rm anom}(\kappa_{SM})=0.
\eeq
We call these curves $SM curves$.
The lighter curves instead correspond to the equality condition
of the two members in eq.(\ref{duesig}). 
Hence, in general, by requiring that no
deviation from the SM rates is observed at  the 95\% CL,
we will be able to
exclude  all the parameters plane but the area 
between the two lighter curves. Note that in some cases
the internal, lighter curve is not shown (see figs. 3-8, 11-16).
Indeed, in these cases the equality in eq.(\ref{duesig}) has just one 
curve as solution.
In the latter cases one will be able to exclude all the 
parameter plane but the area
inside the unique light curve, that of course contains the $SM curve$.

Note also that in many cases we magnified the axis scale 
in order to better distinguish the details of the smaller curves, 
loosing in such a way some portion of the more extended contours, 
that are, however, less relevant in order to set the best limits on 
the anomalous couplings.

In general, the two 95\% CL curves define a $strip$ in the plane,  
some portion of which is quite far from the point (0,0), the  point
where any non-excluded region should be naturally centered around.
These ``far-away" portions can be removed by combining the data 
relative to different
polarizations of the electron beam and/or different sets of kinematical
cuts. For instance, in figure 1, in the plane $(\aa,~\az)$,
by comparing the bounds corresponding
to the $\pe=0,-1,+1$ polarization at $\sqrt{s}=500$~GeV and $\pte>0$,
one can exclude most of the area far from the $(\aa=0,\az=0)$ point.
The remaining three intersection regions which are far from the origin
can be removed by adding the information from figure 5.
Here, the same case of figure 1 is studied with an additional cut 
$\pte>100$~GeV (and $\theta_e<90^o$), and one is left with just 
two intersection regions.
Combining the informations in both figures, the only surviving area
is the one around (0, 0). The extension of this area is of course
strictly connected to the sensibility of the \eaeh process
to the  $\aa$ and $\az$ couplings.

Note that, because of the interference effects with the SM cross
section, the plots are in general asymmetric with respect
to the zero of the $\aa$ and $\az$ couplings. On the other hand,
the CP-odd variables $\aacp$ and $\azcp$ shows modest interference effects,
as discussed in Section 3.

In order to make a more quantitative discussion, it is 
useful to go through the corresponding one-dimensional limits
on the anomalous couplings.

In tables 3-6 we show the 95\% CL limits on each anomalous 
coupling (according to eq.(\ref{duesig})),
when the remaining three couplings are switched off.
Tables 3-4 are relative to the $\sqrt{s}=500$~GeV, 
${\cal L}_{int}=10^2$fb$^{-1}$ case, while tables 5-6 refer
to  $\sqrt{s}=1500$~GeV, ${\cal L}_{int}=10^3$fb$^{-1}$.
We assume $\mh=120$~GeV and the kinematical cuts are the same as 
described above for the corresponding cases in figure 1-16.

In each case, we first show the bounds on the \aah vertex couplings
$(\aa,~\aacp)$ and on the  \zah vertex couplings $(\az,~\azcp)$
(cf. tables 3 and 5). Then we present the corresponding
bounds on the anomalous couplings that are directly associated 
to the operators of the effective lagrangian,
$\d, \db, \bard, \bardb$ (cf. tables 4 and 6).
In each table, three fixed polarization state ($\pe=0,-1,+1$) for 
the electron beam and two different choices for the
$\pte$ cut are presented. 

In general the limits on the different parameters range in the 
interval  
\beq
|\kappa|< 10^{-4} - 10^{-2},
\eeq
also depending on the c.m. energy.
One can see that the choice $\pte>100$~GeV 
improves the limits on the $\az$ and $\azcp$ parameters with
respect to the $\pte>0$ case. The opposite is true for the 
$\aa$ and $\aacp$ couplings, since the \aah vertex mostly contribute to
the low-$\pte$ cross section.

The effect of the $e$-beam polarization is different for 
different parameters.
At $\sqrt{s}=500$~GeV, the $\pe=-1$ polarization improves the limits on
the $\aa$ and $\az$ couplings, especially at large values of $\pte$.
On the other hand, the bounds on the CP violating couplings 
gets weaker with respect to the unpolarized case. 
The opposite happens for the $\pe=1$ polarization. 

Referring to the $\d, \db, \bard$ and $\bardb$ set of bounds,
considerably better limits are obtained for the couplings
$\db$ and $\bardb$, which enter with a coefficient 
$\cotgw \sim 1.8$ in the combinations giving $\aa$ and 
$\aacp$ in eq.(\ref{eq:anomC}).

Concerning the polarization effects, $\pe=1$ improves the limits on
$\db$ and $\bardb$ especially at large $\pte$.
On the other hand,
$\pe=-1$ improves the $\d, \db$, and $\bard$ cases, while worsens
the $\bardb$ case.

A similar discussion applies to the $\sqrt{s}=1.5$~TeV case.
Note that the clear improvement in the set of the coupling bounds
with  respect to the lower c.m energy case is  only slightly due to the
different value of $\sqrt{s}$. The crucial effect comes from the factor
10 of increase in the assumed integrated luminosity
at $\sqrt{s}=1.5$~TeV. 

From the results presented in tables 3-6
we draw the following conclusions:
\begin{itemize}
\item The strongest bounds on the CP-even $\aa$ and CP-odd 
$\aacp$ couplings are at the level of $|\aa| \lta 5\times 10^{-4}$, 
$|\aacp| \lta 2\times 10^{-3}$ at $\sqrt{s}=500$~GeV, and 
$|\aa| \lta 2\times 10^{-4}$, 
$|\aacp| \lta 1\times 10^{-3}$ at $\sqrt{s}=1500$~GeV.
They are obtained in the case when no cut 
on $\pte$ is imposed. 
In both the CP-even and CP-odd cases these bounds are only slightly sensitive 
to the initial $e$-beam polarization.
\item
The strongest limits on the CP-even coupling $d_{\gamma z}$ are 
at the level of
$|\az|\lta 2\times 10^{-3}$ at $\sqrt{s}=500$~GeV, and
$|\az|\lta 3\times 10^{-4}$ at $\sqrt{s}=1500$~GeV, and
are obtained from a left-handed 
polarized electron beam ($\pe=-1$) by imposing a $\pte >100$~GeV cut.
On the other hand,  in the CP-odd case, the right-handed polarized
electron beam ($\pe=1$) gives the best performance,
with bounds at the level of 
$|\azcp|\lta 5\times 10^{-3}$ at $\sqrt{s}=500$~GeV, and
$|\azcp|\lta 1.5\times 10^{-3}$ at $\sqrt{s}=1500$~GeV. 
Indeed, in the latter case, the violation of the strong destructive 
interference between the \aah and \zah terms by the anomalous 
terms compensates the decrease in statistics.
\end{itemize}
The analogous results for the bounds on the CP-even  
$\d,~\db$ and CP-odd $~\bard,~\bardb$
anomalous couplings (see eq.~(\ref{eq:anomC}))
can be outlined as follows:
\begin{itemize}
\item The strongest bounds on the CP-even couplings at $\sqrt{s}=500$~GeV
 are at the level of
$|\d| \lta 6 \times 10^{-4}$, obtained at $\pe=-1$, 
and $|\db| \lta 2.5 \times 10^{-4}$
(with no cut on $\pte$), not depending on the $e$ polarization.
At $\sqrt{s}=1500$~GeV, one has 
$|\d| \lta 1.7 \times 10^{-4}$, obtained at $\pe=-1$ and $\pte >100$~GeV, 
and $|\db| \lta 1 \times 10^{-4}$
(with no cut on $\pte$), not depending on the $e$ polarization.
\item The strongest bounds on the CP-odd couplings at $\sqrt{s}=500$~GeV
are $|\bard| \lta 3 \times 10^{-3}$ and $|\bardb| \lta 1 \times  10^{-3}$, 
which are obtained for $\pe=-1$ and
$\pe=1$, respectively. These bounds are quite 
insensitive to the cuts on $\pte$.
At $\sqrt{s}=1500$~GeV, one has $|\bard| \lta 1.0 \times 10^{-3}$ 
for $\pe=-1$, with $\pte>100$~GeV, and
$|\bardb| \lta 3 \times  10^{-4}$, for $\pe=1$, with $\pte>100$~GeV.
\end{itemize}
Other processes have been studied in the literature that could be able to 
bound the parameters $\d,~\db,~\bard,~\bardb$ at future linear colliders. 
In particular, the processes $e^+ e^- \to HZ$ and $\gamma \gamma \to H$
have been studied for a \epem collider at $\sqrt{s}=1$~TeV and 
with 80 fb$^{-1}$ by Gounaris et al.
From $e^+ e^- \to HZ$, they get 
$|\d| \lta 5 \times 10^{-3}$,
$|\db| \lta 2.5 \times 10^{-3}$, $|\bard| \lta 5 \times 10^{-3}$
and $|\bardb| \lta 2.5 \times 10^{-3}$  \cite{eeHZ}.
The process $\gamma \gamma \to H$ can do a bit better and reach the values
$|\d| \lta 1 \times 10^{-3}$,
$|\db| \lta 3 \times 10^{-4}$, $|\bard| \lta 4 \times 10^{-3}$
and $|\bardb| \lta 1.3 \times 10^{-3}$, assuming a particular photon
energy spectrum \cite{gouna96}.
These analysis assume a precision of the measured 
production rate equal to $1/\sqrt{N}$ (with $N$ the total number of
events), and neglect possible backgrounds.
\noindent
In order to set the comparative potential of our process with respect
to these two processes in bounding the parameters  
$\d,~\db,~\bard,~\bardb$, according to the discussion at the 
end of section 1, we assumed $\sqrt{s}=0.9$~TeV and (conservatively)
a luminosity of  25 fb$^{-1}$ in  \eaeh.
We then neglected any background, and assumed a precision 
equal to $1/\sqrt{N}$. 
In the case $\pe=0$ and $\pte >0$, we get 
$|\d| \lta 5 \times 10^{-4}$,
$|\db| \lta 2 \times 10^{-4}$, $|\bard| \lta 2 \times 10^{-3}$
and $|\bardb| \lta 8 \times 10^{-4}$.
This analysis confirms the excellent potential of the process \eaeh.
Note however that the  $\gamma \gamma \to H$ process has the further 
advantage of isolating the $\aa$ anomalous component
from a possible non-zero $\az$ contribution \cite{ginz}, contrary to the
channel \eaeh, where the effects of the two couplings are superimposed.
\section{Conclusions}

We performed a complete study of the potential of the process
\eaeh for discovering possible \aah and \zah anomalous couplings.
In the effective lagrangian formalism, 
we concentrated on the \xdim operators  that can not be constrained 
through the process $e^+e^- \to WW$.
We included a detailed analysis of all the relevant backgrounds
and the possibility of longitudinal polarization for the electron beam.
We also studied the set of kinematical cuts that optimizes
the signal to background ratio.
We found that the \eaeh gives the possibility to 
improve considerably the constrains on 
the $\d,~\db,~\bard,~\bardb$ couplings that can be obtained by
other processes.
Following the conventions of \cite{9811413},
one can convert these constrains into upper limits of the NP
scale $\Lambda$ that can be explored through 
\eaeh with $\sqrt{s}\simeq 1.5$~TeV and 10$^3$ fb$^{-1}$:
\bea
|\d| \lta 1.7 \times 10^{-4}     
& \to &    |\frac{f_{WW}}{\Lambda^2}| \lta  0.026  ~~ TeV^{-2} \nl 
|\db| \lta 1 \times 10^{-4}     
& \to &    |\frac{f_{BB}}{\Lambda^2}| \lta  0.015  ~~ TeV^{-2} \nl
|\bard| \lta 1 \times 10^{-3}     
& \to &   |\frac{\bar f_{WW}}{\Lambda^2}| \lta  0.15 ~~  TeV^{-2} \nl 
|\bardb| \lta 3 \times 10^{-4}     
& \to &   |\frac{\bar f_{WW}}{\Lambda^2}| \lta  0.046  ~~  TeV^{-2} 
\eea
For $f_i\sim 1$ one can explore energy scales up to about 
6, 8, 2.6 and about 4.5 TeV, respectively. 
At $\sqrt{s} \simeq$ 500 GeV and $10^2$ fb$^{-1}$, the 
corresponding constraints on the couplings are a factor 2 or 3 weaker
than above
(reflecting into energy scales $\Lambda$ lower by a factor 1.4 or 1.7,
respectively),
mainly because of the smaller integrated luminosity assumed.

\section*{Acknowledgements}
E.G. would like to thank C. Di Cairano and A.M. Turiel for useful discussions.
E.G. acknowledges the financial support of the TMR network
project ref. FMRX-CT96-0090 and partial financial support from the 
CICYT project ref. AEN97-1678.
V.I. acknowledges the financial support of the Grant
Center of the St.Petersburg St. University.


\begin{table}
\begin{center}
\begin{tabular}{|r||c|c|}
\hline
$\pe=0 $ & $\pte > 0 $ & $ \pte > 100~\GeV$ 
\\ \hline 
   \hline $\aa  \times 10^{3}$ & $(- 0.44,~0.46)$ & 
                                 $(- 1.1,~1.3)$ 
\\ \hline $\az  \times 10^{3}$ & $(- 5.3,~15)$ & 
                                 $(- 2.7,~5.0)$ 
\\ \hline $\aacp\times 10^{3}$ & $(- 2.0,~2.0)$ & 
                                 $(- 2.7,~2.8)$
\\ \hline $\azcp\times 10^{3}$ & $(- 8.8,~9.0)$ & 
                                 $(- 5.7,~5.9)$
\\ \hline
\hline
$\pe=1 $ & $\pte > 0 $ & $ \pte > 100~\GeV$ 

\\ \hline 
   \hline $\aa  \times 10^{3}$ & $(- 0.46,~0.49)$ & 
                                 $(- 1.5,~3.3)$ 
\\ \hline $\az  \times 10^{3}$ & $(- 13,~7.0)$ & 
                                 $(- 7.3,~3.5)$ 
\\ \hline $\aacp\times 10^{3}$ & $(- 1.9,~1.9)$ & 
                                 $(- 2.2,~2.2)$
\\ \hline $\azcp\times 10^{3}$ & $(- 9.4,~9.5)$ & 
                                 $(- 5.1,~5.1)$
\\ \hline
\hline
$\pe=-1 $ & $\pte > 0 $ & $ \pte > 100~\GeV$ 

\\ \hline 
   \hline $\aa  \times 10^{3}$ & $(- 0.43,~0.45)$ & 
                                 $(- 0.88,~0.88)$ 
\\ \hline $\az  \times 10^{3}$ & $(- 3.1,~4.1)$& 
                                 $(- 1.7,~1.7)$ 
\\ \hline $\aacp\times 10^{3}$ & $(- 2.0,~2.0)$ & 
                                 $(- 2.9,~3.1)$
\\ \hline $\azcp\times 10^{3}$ & $(- 10,~11)$ & 
                                 $(- 6.5,~6.9)$
\\ \hline
\end{tabular}
\caption[]{Bounds on the CP-even $\aa,~\az$
and CP-odd $\aacp$ $\azcp$ anomalous couplings, 
for $\sqrt{S}=500~\GeV$, $e$-beam polarizations $\pe=0,1,-1$, and 
for $\pte>0$ and $\pte> 100~\GeV$. In the case $\pte> 100~\GeV$, 
a cut on the final electron angle $\theta_e < 90^o$ is applied (see text). }
\label{tab500_zero}
\end{center}
\end{table}
\begin{table}
\begin{center}
\begin{tabular}{|r||c|c|}
\hline
$\pe=0 $ & $\pte > 0 $ & $ \pte > 100~\GeV$ 
\\ \hline 
   \hline $\d~   \times  10^{3}$ &  $(- 0.73,~0.76)$ & 
                                    $(- 1.2,~1.3)$ 
\\ \hline $\db  \times  10^{3}$ &   $(- 0.25,~0.26)$ & 
                                    $(- 0.70,~3.3)$ 
\\ \hline $\bard~\times  10^{3}$ &  $(- 3.2,~3.3)$ & 
                                    $(- 3.6,~3.7)$
\\ \hline $\bardb\times 10^{3}$ &   $(- 1.1,~1.1)$ & 
                                    $(- 1.5,~1.5)$
\\ \hline
\hline
$\pe=1 $ & $\pte > 0 $ & $ \pte > 100~\GeV$ 
\\ \hline 
   \hline $d~  \times 10^{3}$ &   $(- 0.89,~0.94)$ & 
                                  $(- 10,~25)$ 
\\ \hline $\db  \times 10^{3}$ &  $(- 0.24,~0.26)$ & 
                                  $(- 0.65,~1.4)$ 
\\ \hline $\bard~\times 10^{3}$ & $(- 3.9,~3.9)$ & 
                                  $(- 15,~15)$
\\ \hline $\bardb\times 10^{3}$ & $(- 0.97,~0.96)$ & 
                                  $(- 0.97,~0.97)$
\\ \hline
\hline
$\pe=-1 $ & $\pte > 0 $ & $ \pte > 100~\GeV$ 
\\ \hline 
   \hline $\d~  \times 10^{3}$ &  $(- 0.63,~0.66)$ & 
                                  $(- 0.83,~0.83)$ 
\\ \hline $\db  \times 10^{3}$ &  $(- 0.25,~0.26)$ & 
                                  $(- 0.67,~0.66)$ 
\\ \hline $\bard~\times 10^{3}$ & $(- 3.1,~3.2)$ & 
                                  $(- 2.9,~3.1)$
\\ \hline $\bardb\times 10^{3}$ & $(- 1.2,~1.2)$ & 
                                  $(- 2.3,~2.4)$
\\ \hline
\end{tabular}
\caption[]{Bounds on the CP-even $\d,~\db$
and CP-odd $\bard,~\bardb$ anomalous couplings, 
for $\sqrt{S}=500~\GeV$, $e$-beam polarizations $\pe=0,1,-1$, 
and for $\pte>0$ and $\pte> 100~\GeV$.
In the case $\pte> 100~\GeV$, 
a cut on the final electron angle $\theta_e < 90^o$ is applied (see text). }
\label{tabdb500_zero}
\end{center}
\end{table}

\begin{table}
\begin{center}
\begin{tabular}{|r||c|c|}
\hline
$\pe=0 $ & $\pte > 0 $ & $ \pte > 100~\GeV$ 
\\ \hline 
   \hline $\aa  \times 10^{3}$ & $(- 0.17,~0.17)$ & 
                                 $(- 0.25,~0.26)$ 
\\ \hline $\az  \times 10^{3}$ & $(- 0.98,~1.5)$ & 
                                 $(- 0.51,~0.58)$ 
\\ \hline $\aacp\times 10^{3}$ & $(- 1.0,~1.0)$ & 
                                 $(- 0.96,~0.98)$
\\ \hline $\azcp\times 10^{3}$ & $(- 2.6,~2.6)$ & 
                                 $(- 1.8,~1.8)$
\\ \hline
\hline
$\pe=1 $ & $\pte > 0 $ & $ \pte > 100~\GeV$ 
\\ \hline 
   \hline $\aa  \times 10^{3}$ & $(-0.21 ,~0.22 )$ & 
                                 $(-0.44 ,~1.2 )$ 
\\ \hline $\az  \times 10^{3}$ & $(-3.8 ,~2.0 )$ & 
                                 $(-2.1 ,~1.0 )$ 
\\ \hline $\aacp\times 10^{3}$ & $(-0.96 ,~0.99 )$ & 
                                 $(-0.73 ,~0.71 )$
\\ \hline $\azcp\times 10^{3}$ & $(-2.7 ,~2.7 )$ & 
                                 $(-1.5 ,~1.5 )$
\\ \hline
\hline
$\pe=-1 $ & $\pte > 0 $ & $ \pte > 100~\GeV$
\\ \hline 
   \hline $\aa  \times 10^{3}$ & $(-0.15 ,~0.15 )$ & 
                                 $(-0.18 ,~0.18 )$ 
\\ \hline $\az  \times 10^{3}$ & $(-0.53 ,~0.57 )$ & 
                                 $(-0.33 ,~0.33 )$ 
\\ \hline $\aacp\times 10^{3}$ & $(-1.1 ,~1.1 )$ & 
                                 $(-1.1 ,~1.1 )$
\\ \hline $\azcp\times 10^{3}$ & $(-2.6 ,~2.6 )$ & 
                                 $(-1.9 ,~1.9 )$
\\ \hline
\end{tabular}
\caption[]{Bounds on the CP-even $\aa,~\az$
and CP-odd $\aacp$ $\azcp$ anomalous couplings, 
for $\sqrt{S}=1.5$TeV, $e$-beam polarizations $\pe=0,1,-1$, and for  
$\pte>0$ and $\pte> 100$~GeV (no restriction on the final
electron angle).}
\label{tab1500_zero}
\end{center}
\end{table}
\begin{table}
\begin{center}
\begin{tabular}{|r||c|c|}
\hline
$\pe=0 $ & $\pte > 0 $ & $ \pte > 100~\GeV$
\\ \hline 
   \hline $\d~   \times  10^{3}$ &  $(-0.24 ,~0.25 )$ & 
                                    $(-0.25 ,~0.25 )$ 
\\ \hline $\db  \times  10^{3}$ &   $(-0.10 ,~0.10 )$ & 
                                    $(-0.17 ,~0.21 )$ 
\\ \hline $\bard~\times  10^{3}$ &  $(-1.5 ,~1.5 )$ & 
                                    $(-1.2 ,~1.2 )$
\\ \hline $\bardb\times 10^{3}$ &   $(-5.6 ,~5.6 )$ & 
                                    $(-0.52 ,~0.53 )$
\\ \hline
\hline
$\pe=1 $ & $\pte > 0 $ & $ \pte > 100~\GeV$
\\ \hline 
   \hline $\d~   \times  10^{3}$ &  $(-0.4 ,~0.4 )$ & 
                                    $(-3.3 ,~17 )$ 
\\ \hline $\db  \times  10^{3}$ &   $(-0.11 ,~0.12 )$ & 
                                    $(-0.19 ,~0.49 )$ 
\\ \hline $\bard~\times  10^{3}$ &  $(-2.5 ,~2.7 )$ & 
                                    $(-9.3 ,~8.1 )$
\\ \hline $\bardb\times 10^{3}$ &   $(-4.5 ,~4.7 )$ & 
                                    $(-0.31 ,~0.31 )$
\\ \hline
\hline
$\pe=-1 $ & $\pte > 0 $ & $ \pte > 100~\GeV$ 
\\ \hline 
   \hline $\d~   \times  10^{3}$ &  $(-0.18 ,~0.18 )$ & 
                                    $(-0.16 ,~0.17)$ 
\\ \hline $\db  \times  10^{3}$ &   $(-0.093 ,~0.094 )$ & 
                                    $(-0.14 ,~0.14 )$ 
\\ \hline $\bard~\times  10^{3}$ &  $(-1.2 ,~1.2 )$ & 
                                    $(-0.96 ,~0.98 )$
\\ \hline $\bardb\times 10^{3}$ &   $(-0.69 ,~0.71 )$ & 
                                    $(-0.88 ,~0.90 )$
\\ \hline
\end{tabular}
\caption[]{Bounds on the CP-even $\d,~\db$
and CP-odd $\bard,~\bardb$ anomalous couplings, 
for $\sqrt{S}=1500$~GeV, $e$-beam polarizations $\pe=0,1,-1$, 
and for $\pte>0$ and $\pte> 100$~GeV (no restriction on the final
electron angle).}
\label{tabdb1500_zero}
\end{center}
\end{table}

\newpage
\begin{figure}[t]
\centerline{\epsfig{figure=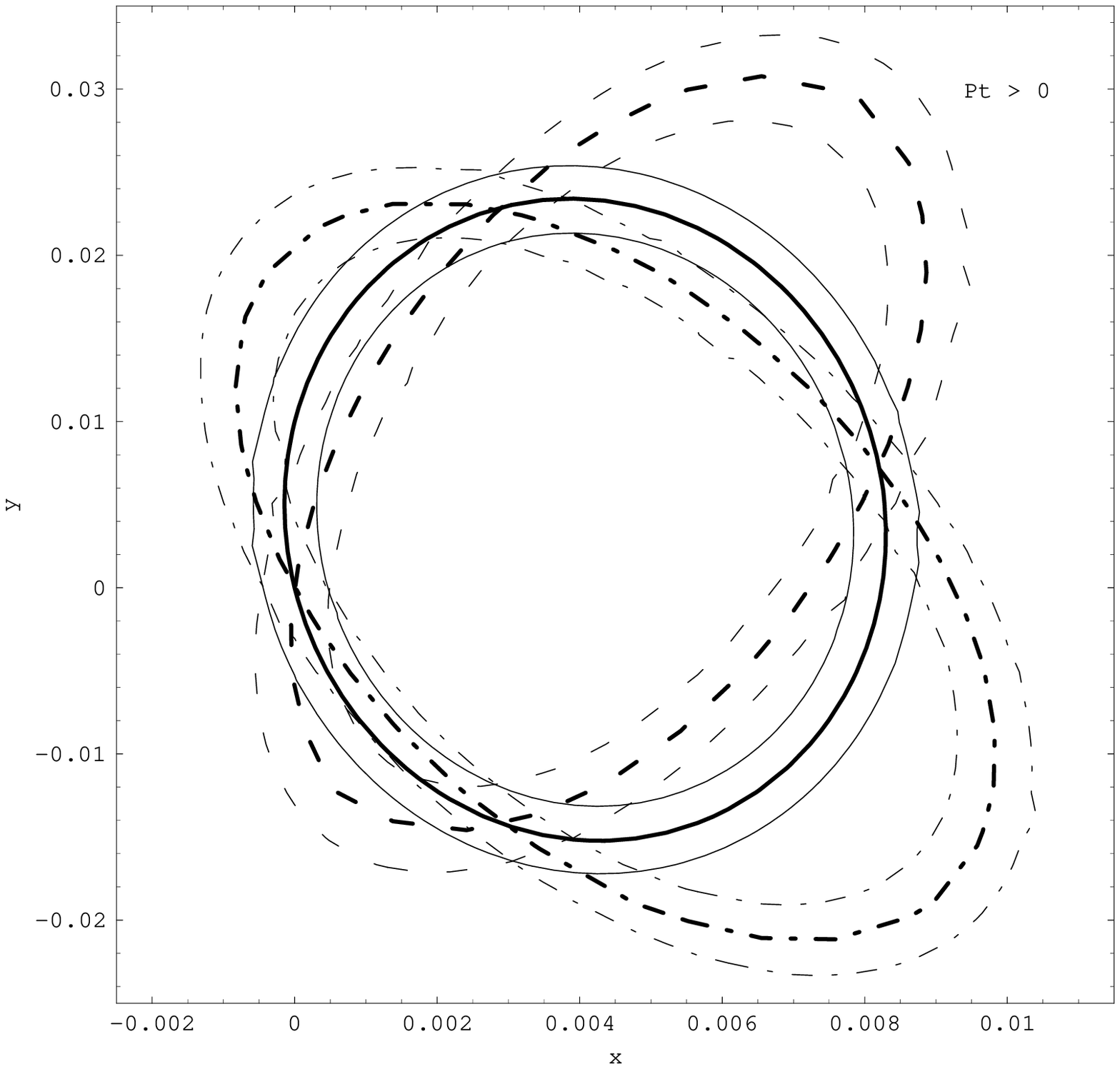,height=15cm,angle=0}}
\caption{
{\small
Contour plots for $\sqrt{S}=500$~GeV with $x=\aa$ and $y=\az$ and
with a cut $\pte >0$ (no restriction on the final electron 
angle $\theta_e$).
The continuous, dashed and dashed-dot lines correspond 
to $P_e=0,1,-1$ e-beam polarization, respectively. 
The darker curves correspond to the Standard Model solution.
}}
\label{naa_az_pt0}  
\end{figure}
\begin{figure}[t]
\centerline{\epsfig{figure=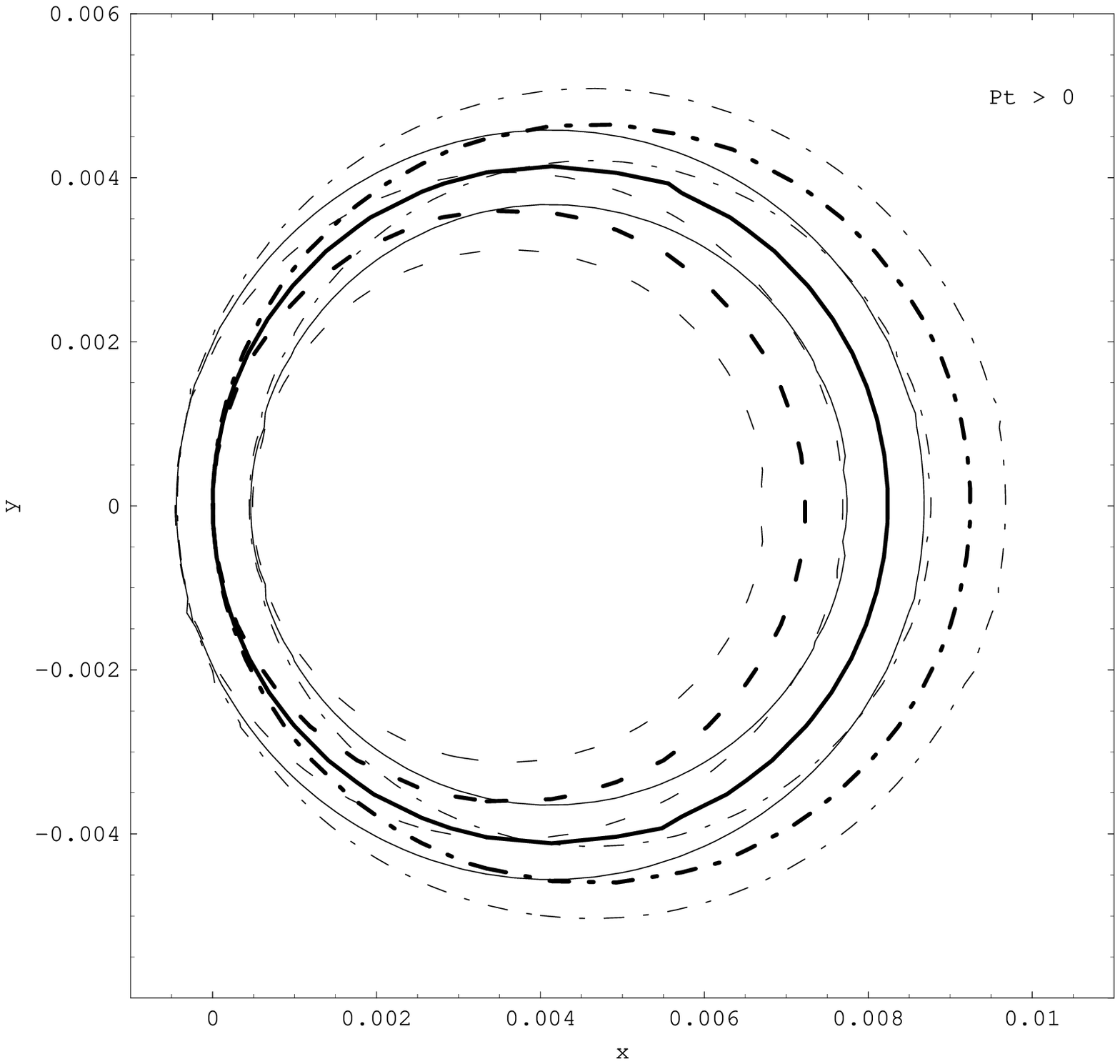,height=15cm,angle=0}}
\caption{
{\small Contour plots as in figure 1 
with  $x=\aa$ and $y=\aacp$. }}
\label{naa_aacp_pt0}  
\end{figure}
\begin{figure}[t]
\centerline{\epsfig{figure=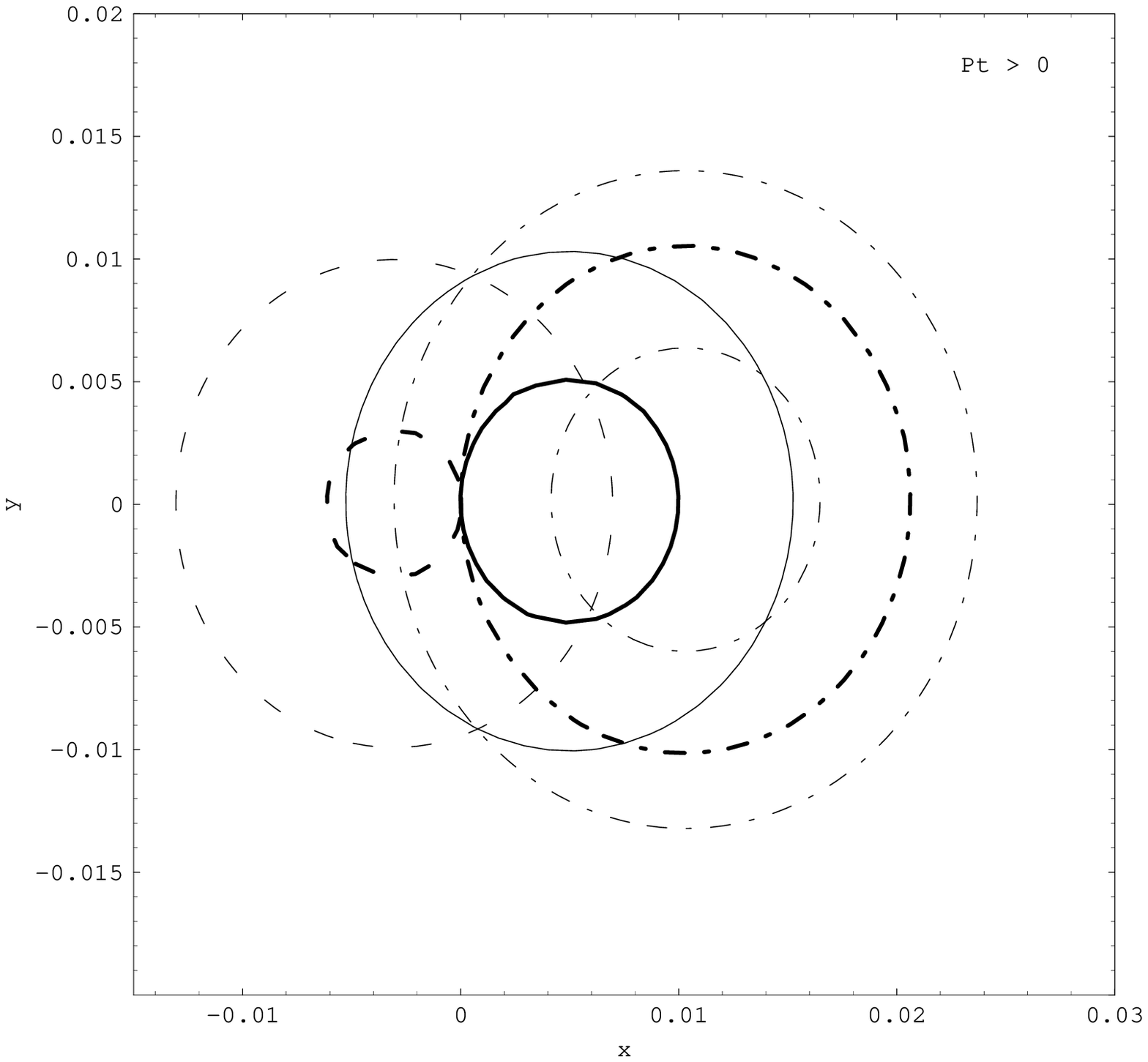,height=15cm,angle=0}}
\caption{
{\small Contour plots as in figure 1 
with $x=\az$ and $y=\azcp$. }}
\label{naz_azcp_pt0}  
\end{figure}
\begin{figure}[t]
\centerline{\epsfig{figure=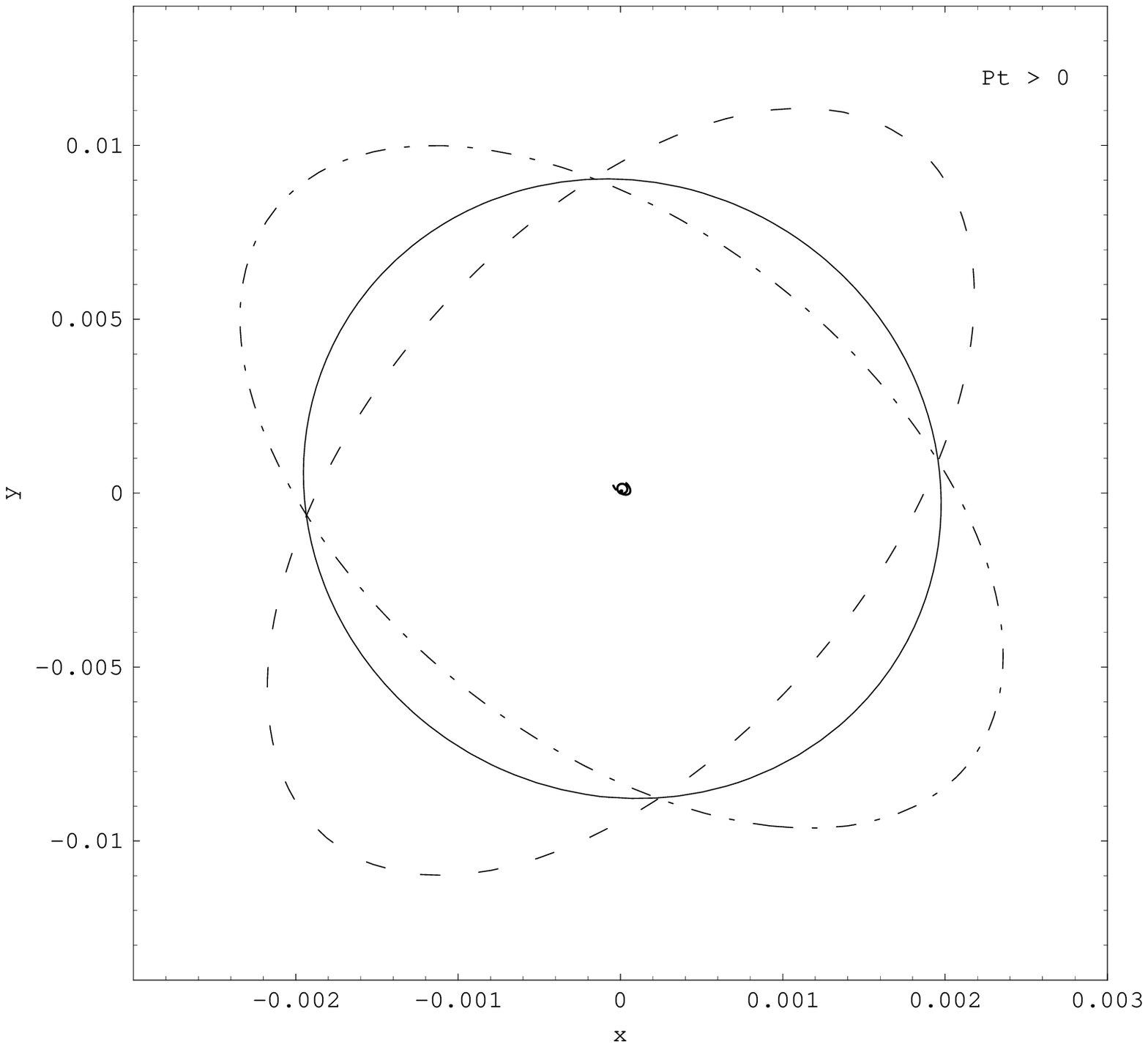,height=15cm,angle=0}}
\caption{
{\small Contour plots as in figure 1 
with $x=\aacp$ and $y=\azcp$. }}
\label{naacp_azcp_pt0}  
\end{figure}
\begin{figure}[t]
\centerline{\epsfig{figure=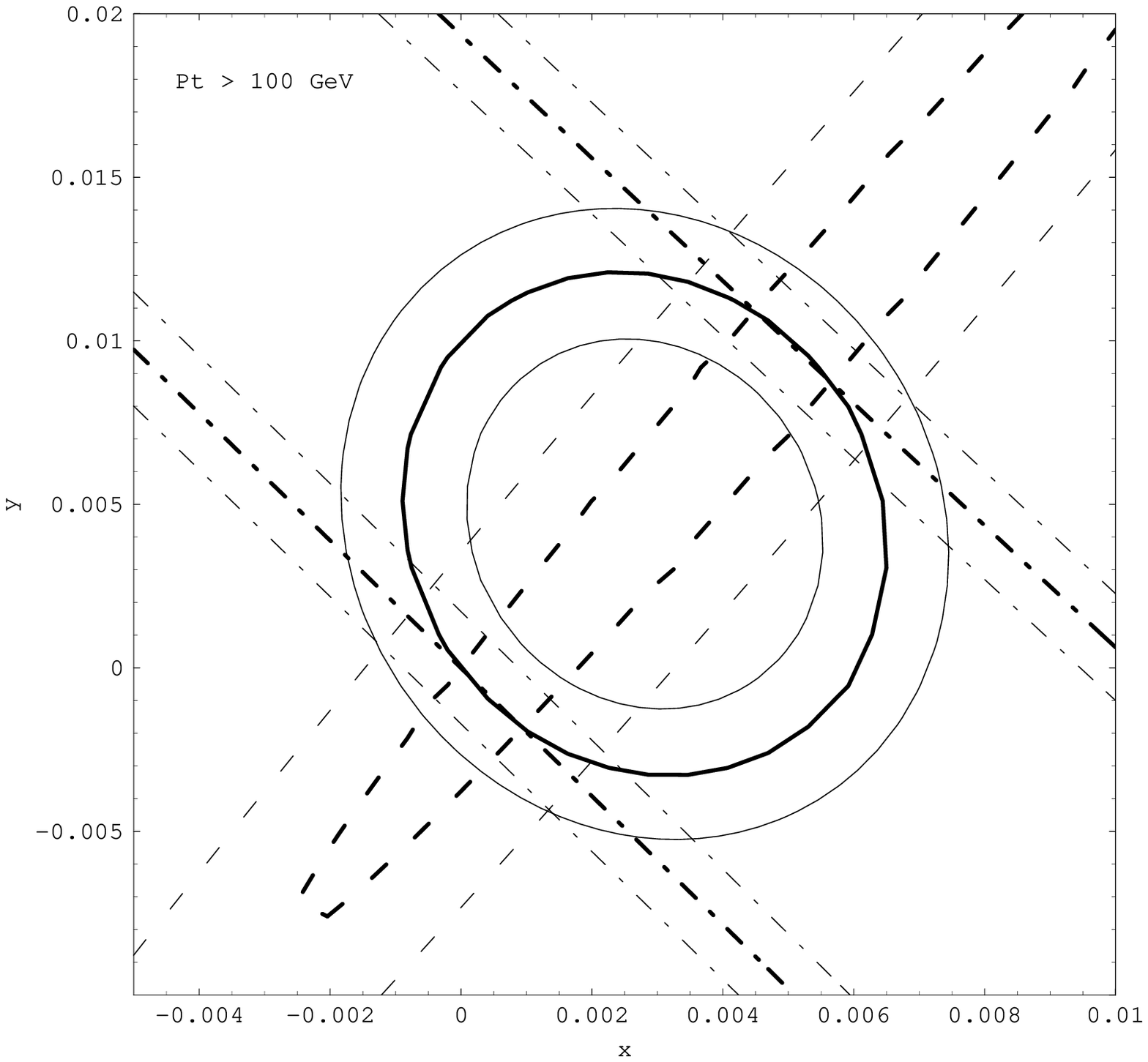,height=15cm,angle=0}}
\caption{
{\small
Contour plots as in figure 1 with $x=\aa$ and $y=\az$,
where the cuts $\pte >100$~GeV and $\theta_e <90^{0}$ 
on the final electron have been applied.}}
\label{naa_az_pt100}  
\end{figure}
\begin{figure}[t]
\centerline{\epsfig{figure=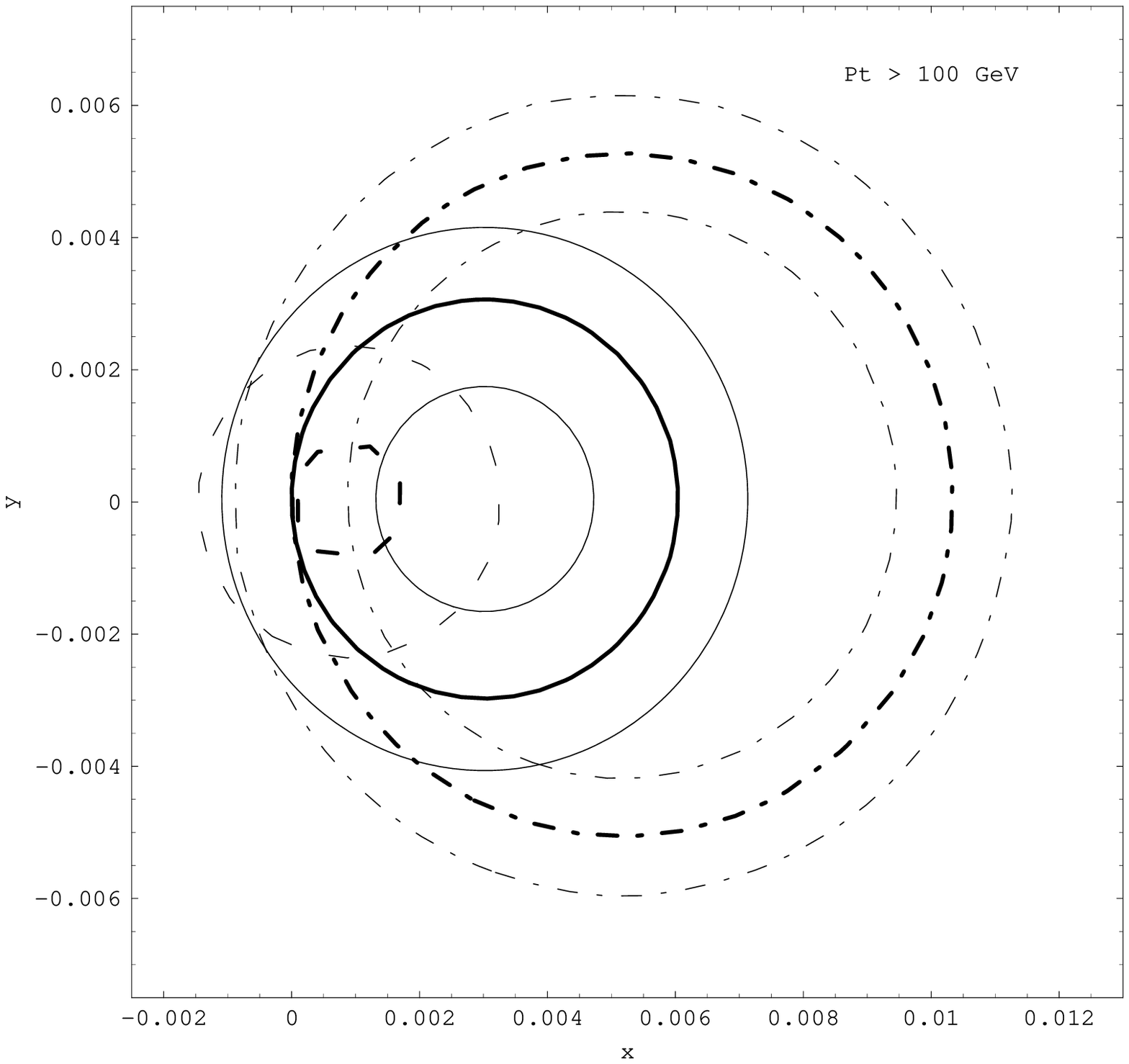,height=15cm,angle=0}}
\caption{
{\small Contour plots as in figure 5 
with  $x=\aa$ and $y=\aacp$. }}
\label{naa_aacp_pt100}  
\end{figure}
\begin{figure}[t]
\centerline{\epsfig{figure=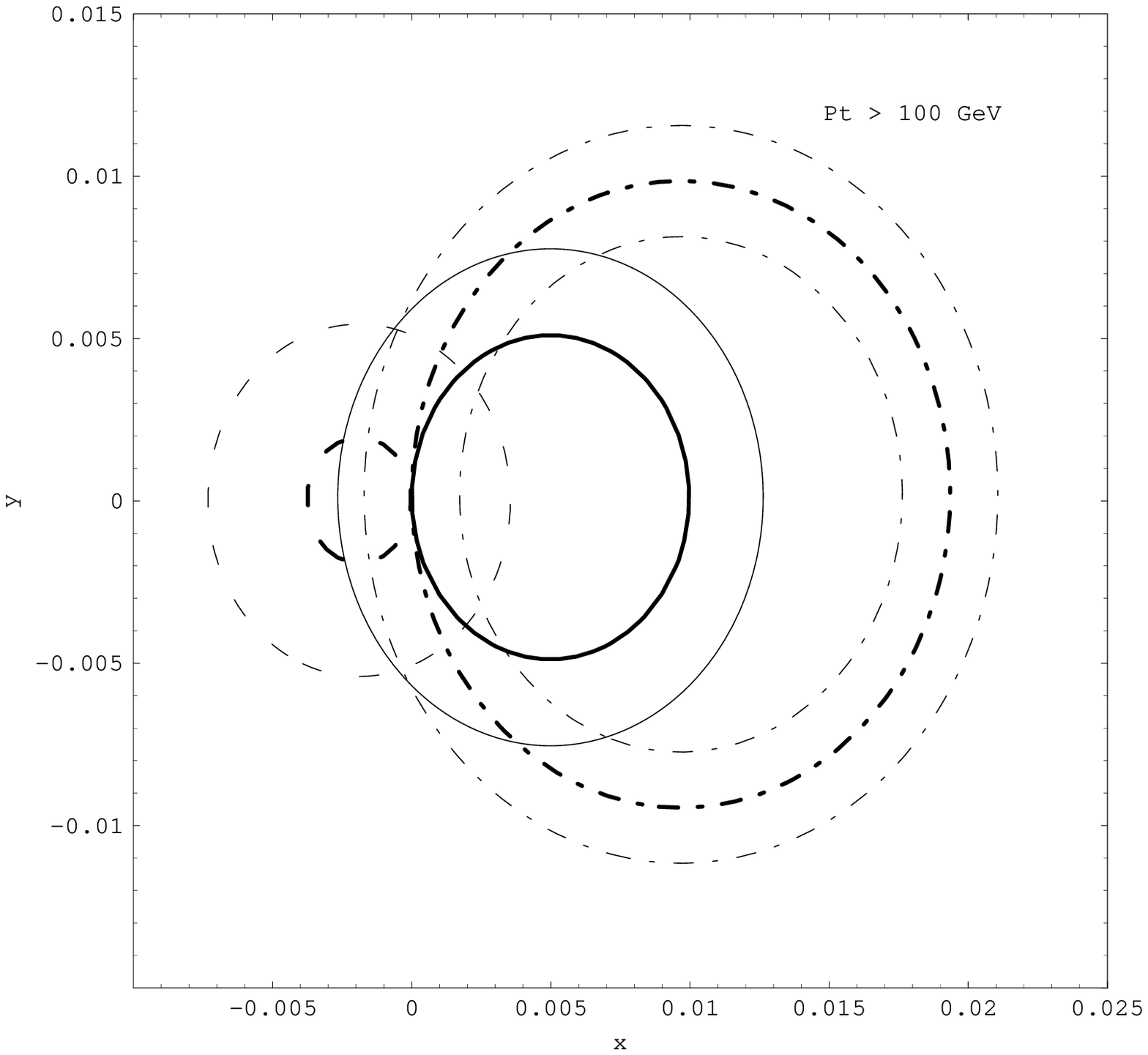,height=15cm,angle=0}}
\caption{
{\small Contour plots as in figure 5 
with $x=\az$ and $y=\azcp$. }}
\label{naz_azcp_pt100}  
\end{figure}
\begin{figure}[t]
\centerline{\epsfig{figure=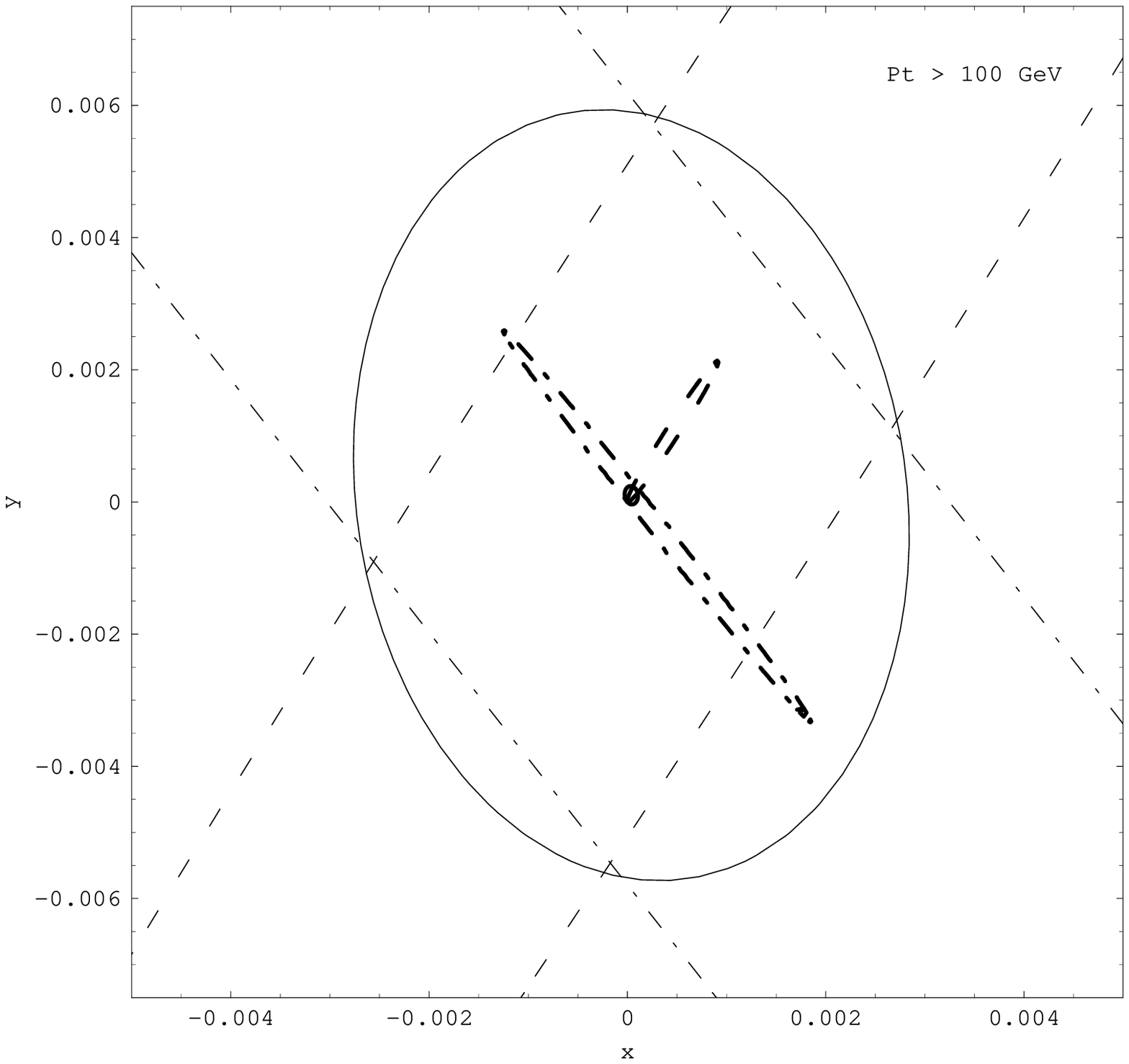,height=15cm,angle=0}}
\caption{
{\small Contour plots as in figure 5
with $x=\aacp$ and $y=\azcp$. }}
\label{naacp_azcp_pt100}  
\end{figure}
\begin{figure}[t]
\centerline{\epsfig{figure=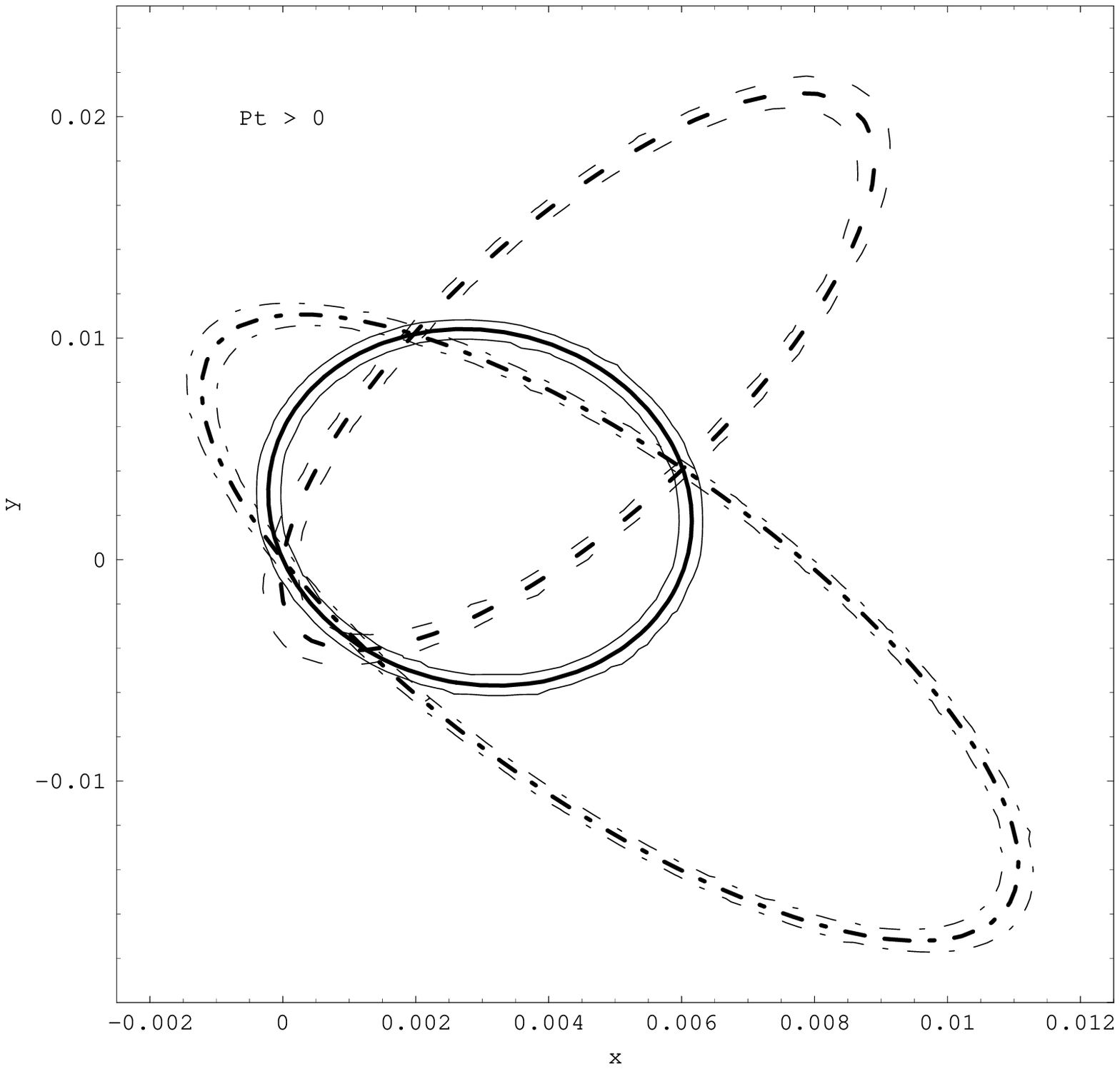,height=15cm,angle=0}}
\caption{
{\small
Contour plots for $\sqrt{S}=1500$~GeV with $x=\aa$ and $y=\az$ and
with a cut $\pte >0$ (no restriction on the final electron 
angle $\theta_e$).
The continuous, dashed and dashed-dot lines correspond 
to $P_e=0,1,-1$ e-beam polarization, respectively. 
The darker curves correspond to the Standard Model solution.
}}
\label{n15aa_az_pt0}  
\end{figure}
\begin{figure}[t]
\centerline{\epsfig{figure=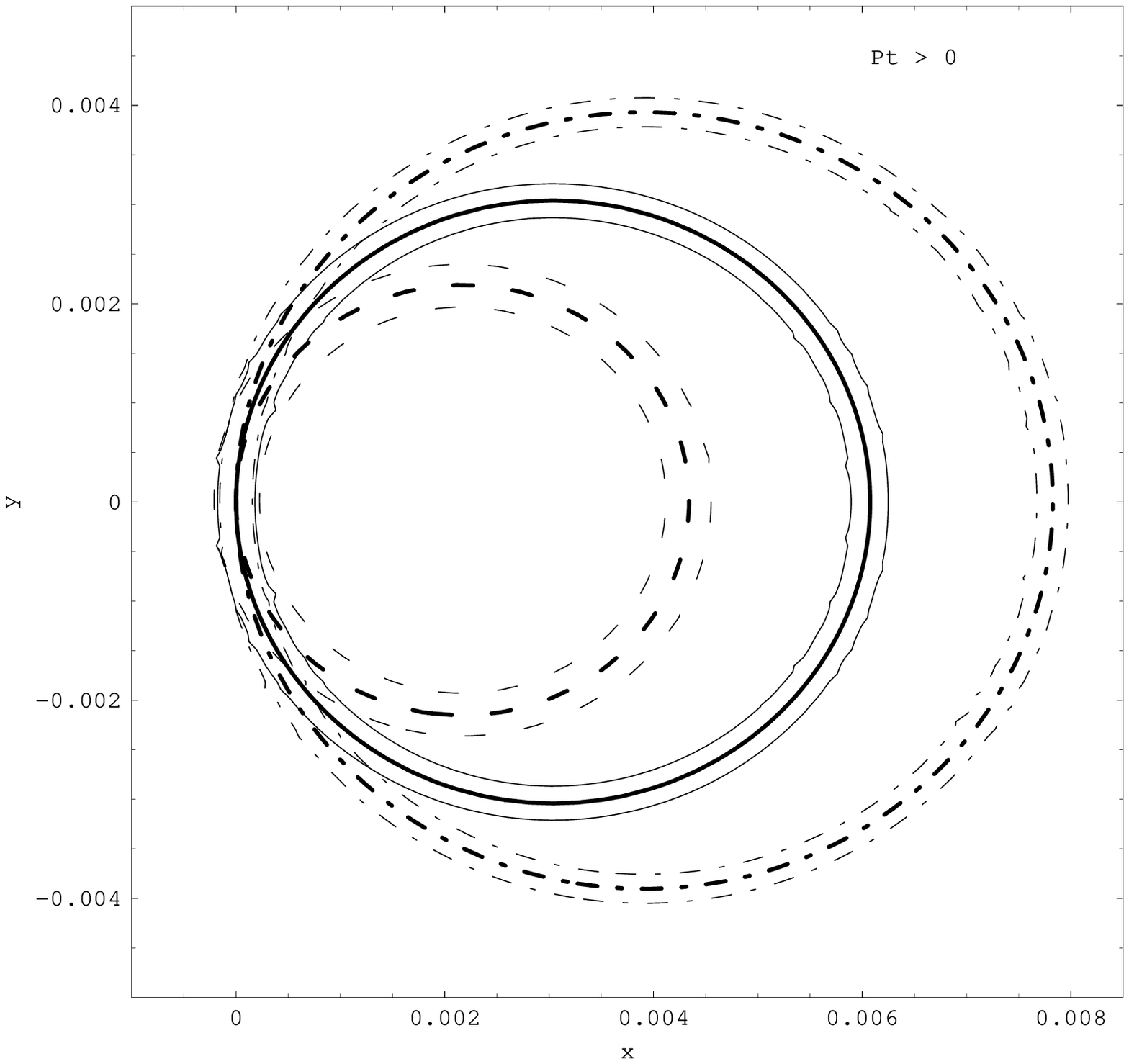,height=15cm,angle=0}}
\caption{
{\small Contour plots as in figure 9 
with  $x=\aa$ and $y=\aacp$ }}
\label{n15aa_aacp_pt0}  
\end{figure}
\begin{figure}[t]
\centerline{\epsfig{figure=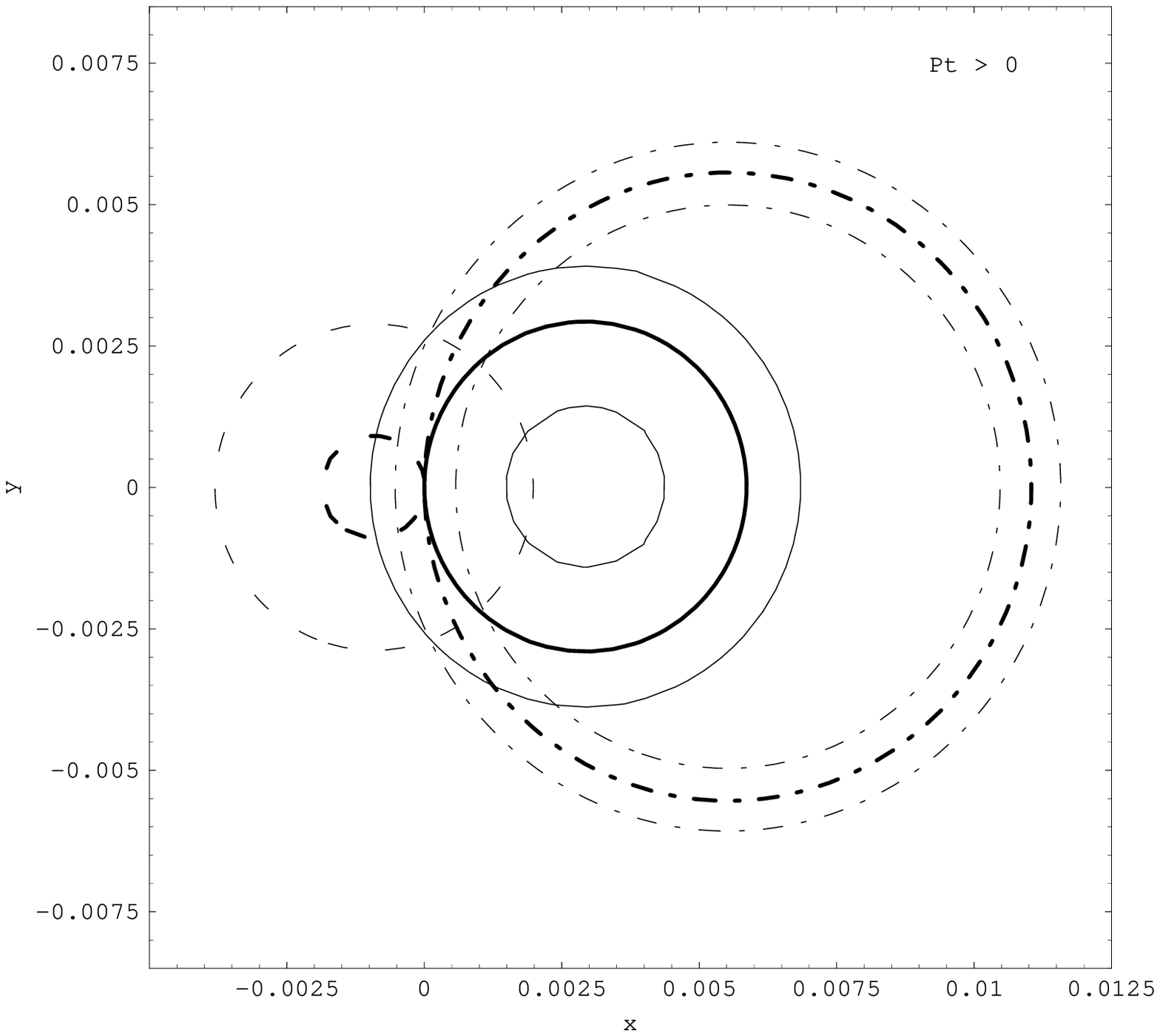,height=15cm,angle=0}}
\caption{
{\small Contour plots as in figure 9 
with $x=\az$ and $y=\azcp$ }}
\label{n15az_azcp_pt0}  
\end{figure}
\begin{figure}[t]
\centerline{\epsfig{figure=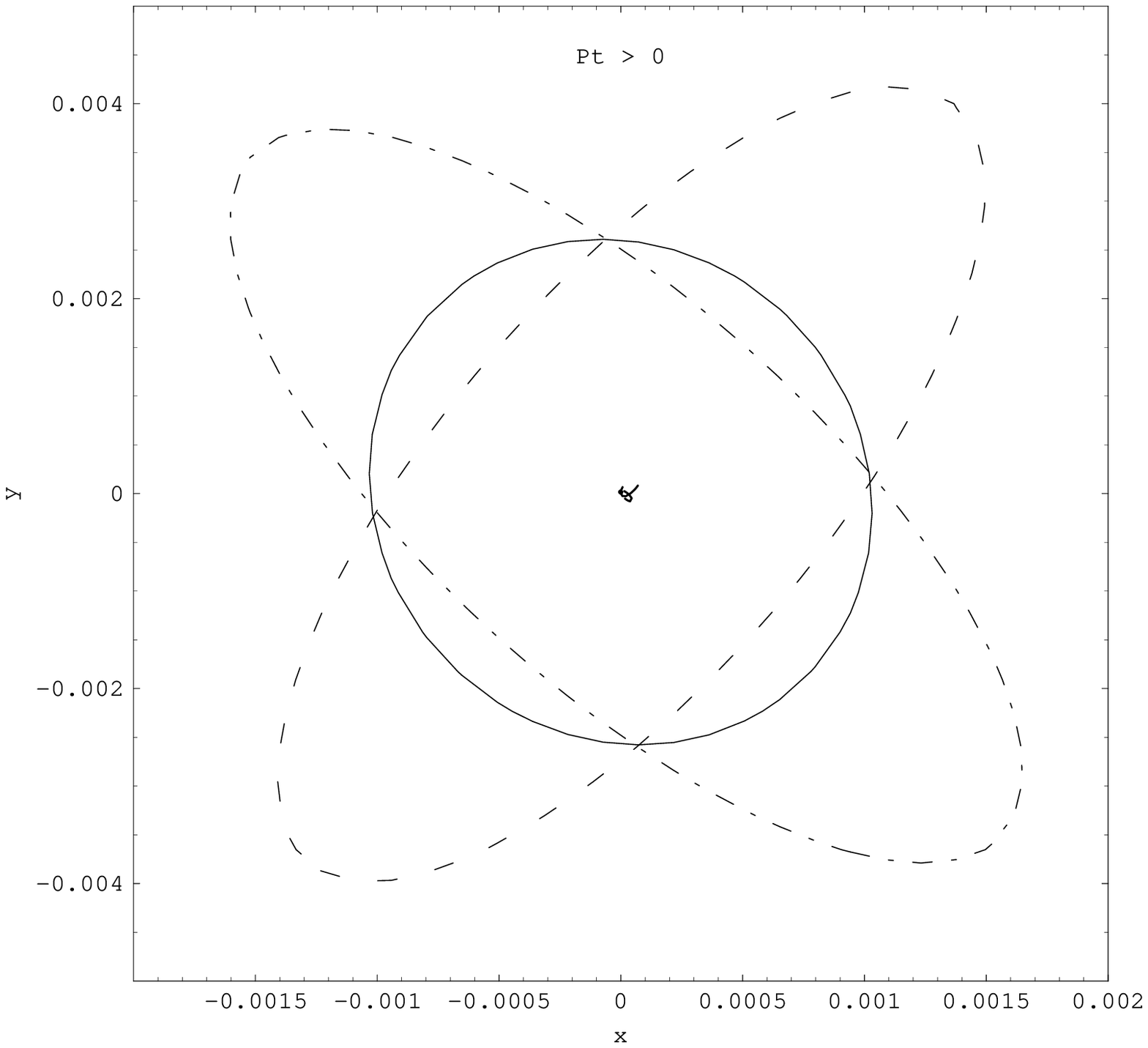,height=15cm,angle=0}}
\caption{
{\small Contour plots as in figure 9 
with $x=\aacp$ and $y=\azcp$ }}
\label{n15aacp_azcp_pt0}  
\end{figure}
\begin{figure}[t]
\centerline{\epsfig{figure=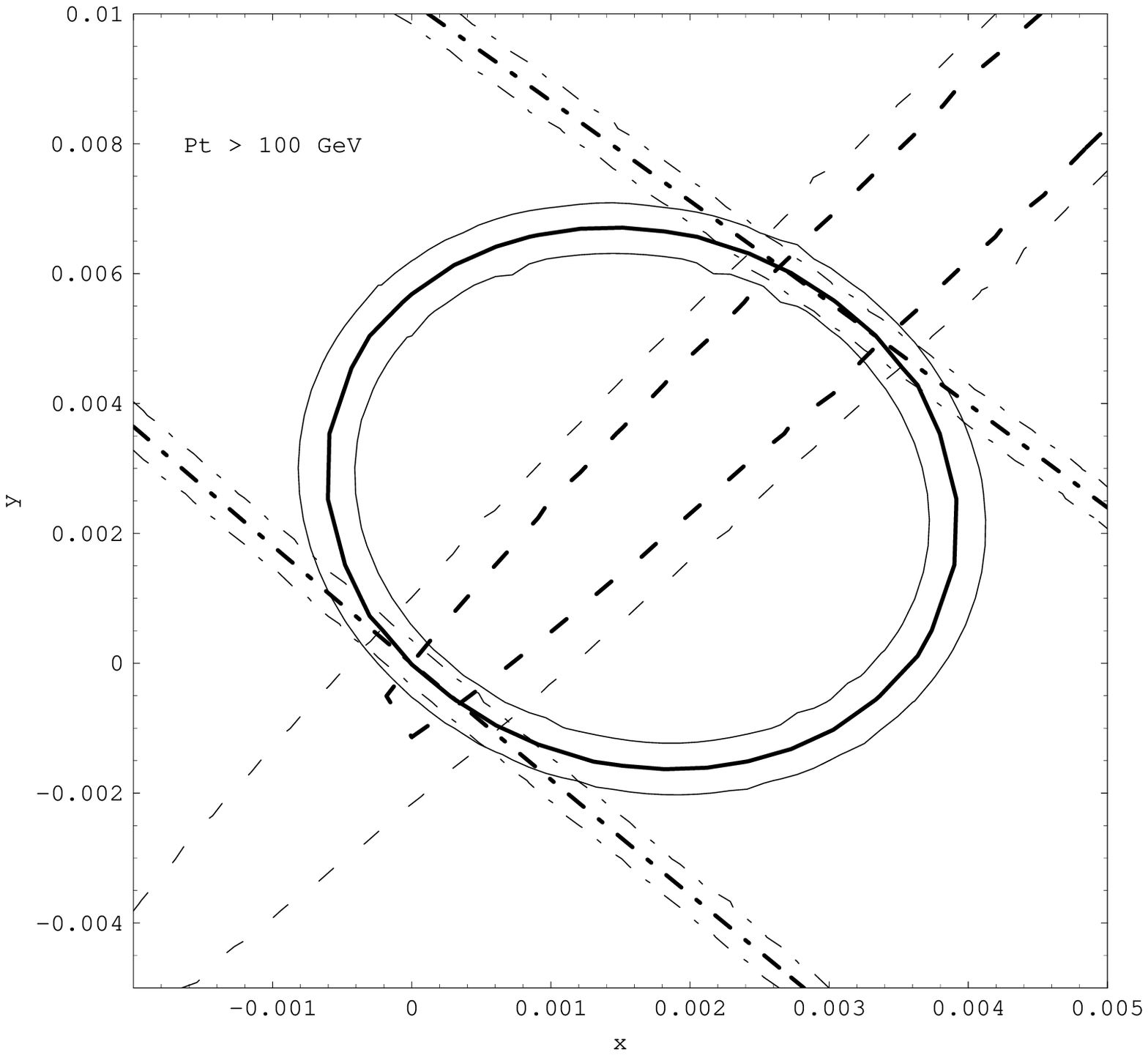,height=15cm,angle=0}}
\caption{
{\small Contour plots as in figure 9 with $x=\aa$ 
and $y=\az$, with $\pte >100$~GeV (no restriction on the final electron 
angle $\theta_e$).}}
\label{n15aa_az_pt100}  
\end{figure}
\begin{figure}[t]
\centerline{\epsfig{figure=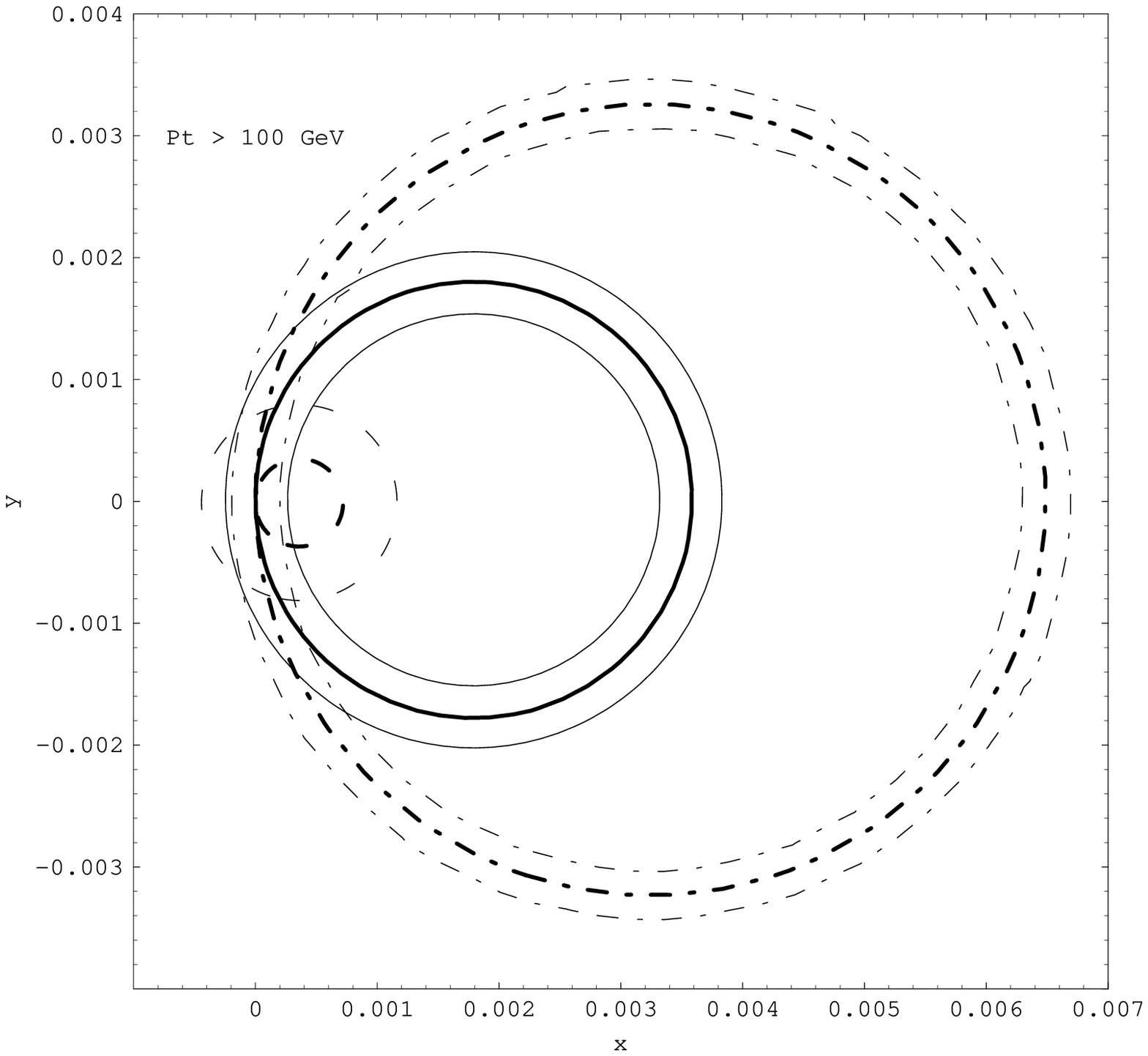,height=15cm,angle=0}}
\caption{
{\small Contour plots as in figure 13
with  $x=\aa$ and $y=\aacp$.}}
\label{n15aa_aacp_pt100}  
\end{figure}
\begin{figure}[t]
\centerline{\epsfig{figure=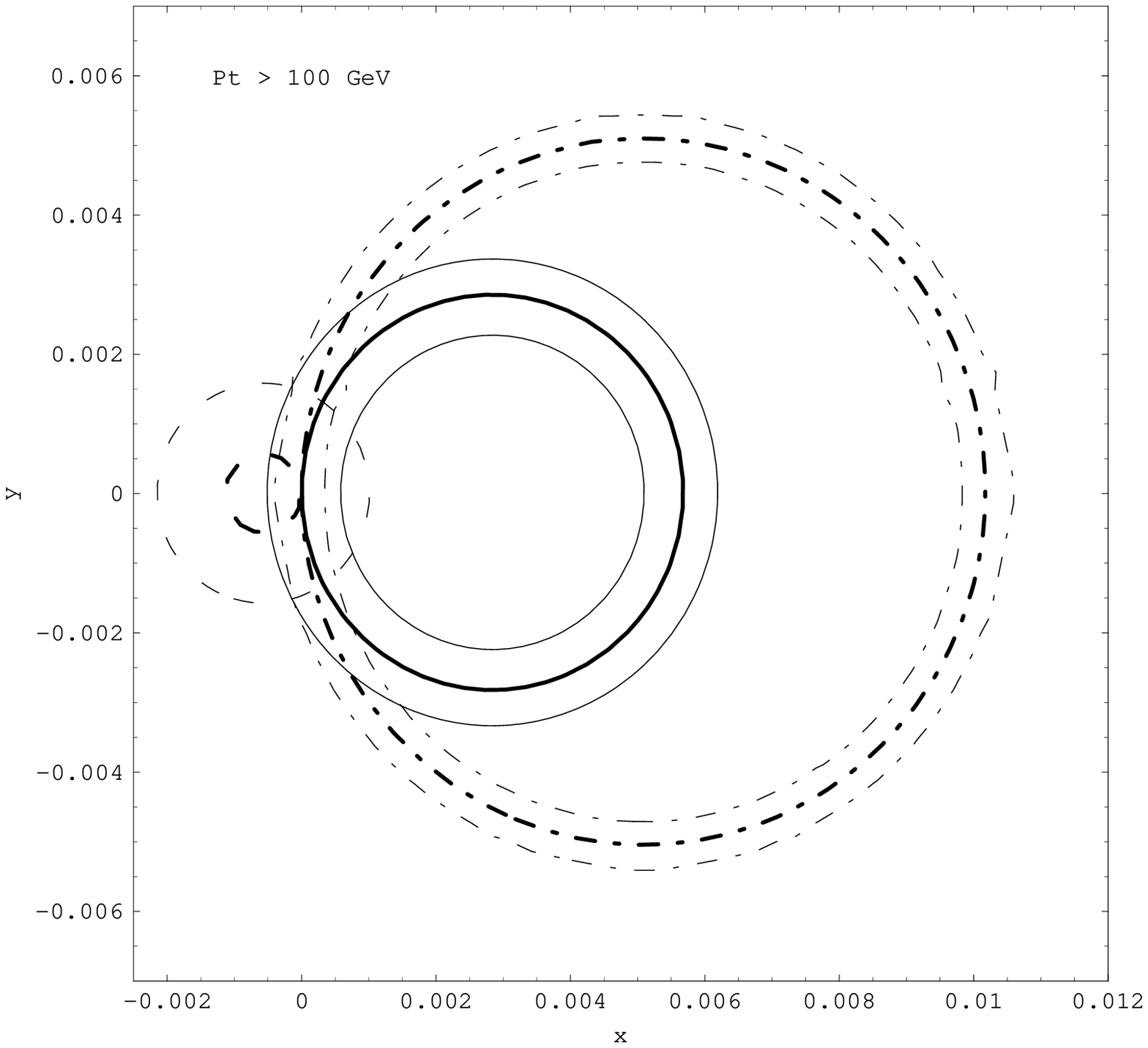,height=15cm,angle=0}}
\caption{
{\small Contour plots as in figure 13
with $x=\az$ and $y=\azcp$.}}
\label{n15az_azcp_pt100}  
\end{figure}
\begin{figure}[t]
\centerline{\epsfig{figure=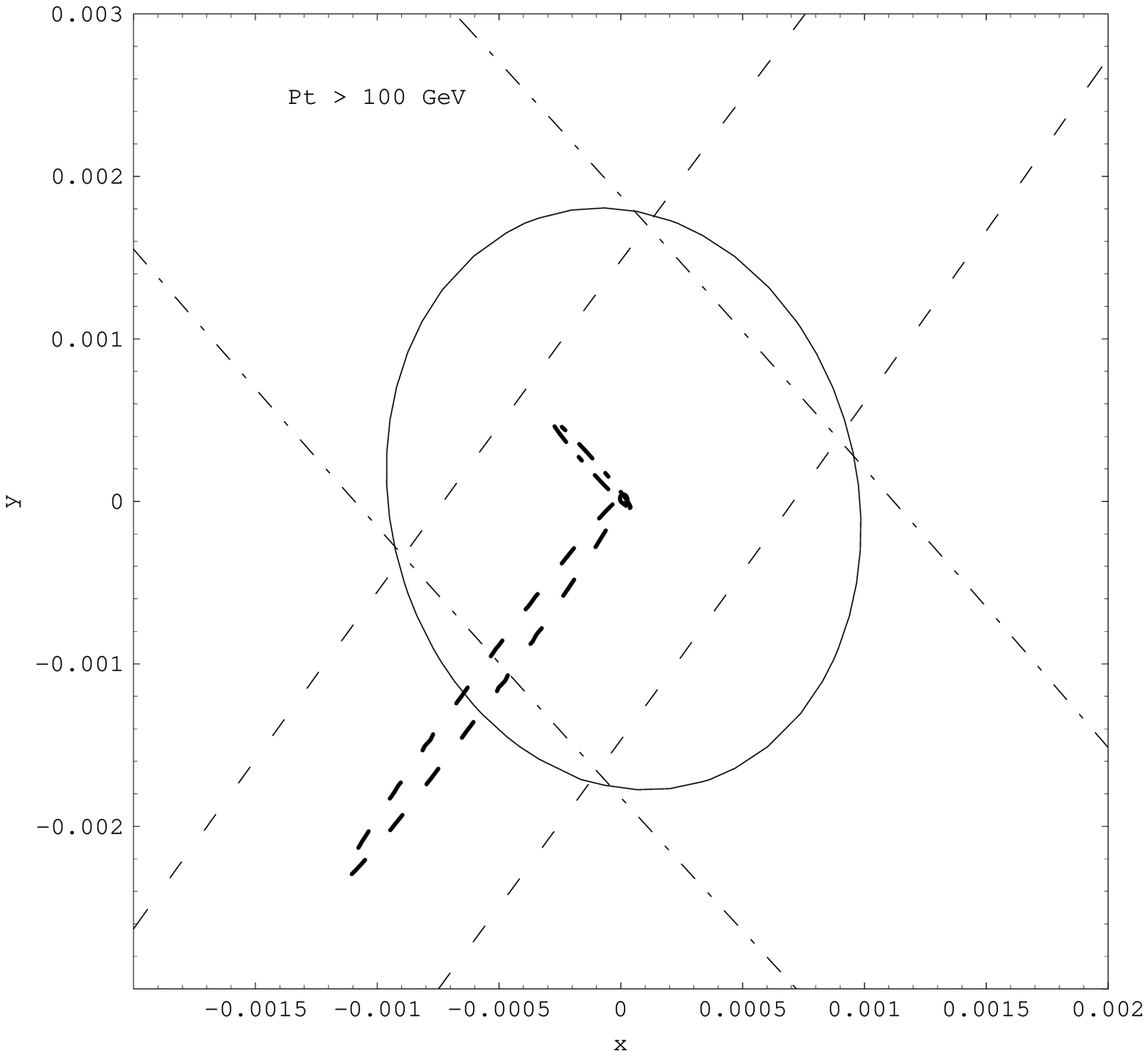,height=15cm,angle=0}}
\caption{
{\small Contour plots as in figure 13
with $x=\aacp$ and $y=\azcp$.}}
\label{n15aacp_azcp_pt100}  
\end{figure}
\end{document}